\documentclass[pra,reprint,amsmath,amssymb,nofootinbib]{revtex4-1}

\usepackage{amsmath}
\usepackage{physics}
\usepackage{graphicx}
\usepackage{dcolumn}
\usepackage[caption=false]{subfig}
\usepackage{bm}
\usepackage{xcolor}
\usepackage{hyperref}

\def\ssB{{\scriptscriptstyle B}}
\def\ssE{{\scriptscriptstyle E}}
\def\ssF{{\scriptscriptstyle F}}
\def\ssC{{\scriptscriptstyle C}}
\def\bfv{{\bf v}}
\def\bfE{{\bf E}}
\def\bfA{{\bf A}}
\def\bfB{{\bf B}}
\def\bfr{{\bf r}}
\def\bfx{{\bf x}}
\def\cM{\mathcal{M}}
\def\cO{\mathcal{O}}
\def\ba{\begin{eqnarray}}
\def\ea{\end{eqnarray}}
\def\be{\begin{equation}}
\def\ee{\end{equation}}
\def\bea{\begin{eqnarray}}
\def\eea{\end{eqnarray}}
\def\exd{{\hbox{d}}}

\def\Psibar{\overline{\Psi}}
\newbox\charbox
\newbox\slabox
\def\slsh#1{{      
        \setbox\charbox=\hbox{$#1$}
        \setbox\slabox=\hbox{$/$}
        \dimen\charbox=\ht\slabox
        \advance\dimen\charbox by -\dp\slabox
        \advance\dimen\charbox by -\ht\charbox
        \advance\dimen\charbox by \dp\charbox
        \divide\dimen\charbox by 2
        \raise-\dimen\charbox\hbox to \wd\charbox{\hss/\hss}
        \llap{$#1$}
}}

\def\Dsl{\slsh{D}}
\def\nn{\nonumber}

\draft 

\begin{document}

\title{Reduced Theoretical Error for \texorpdfstring{${}^4\hbox{\bf He}^+$}{4He+} Spectroscopy}

\author{C.P.~Burgess}
\email{cburgess@perimeterinstitute.ca} 
\author{P.~Hayman}%
\email{haymanpf@mcmaster.ca}
\author{Markus Rummel}
\email{rummelm@mcmaster.ca}
\author{L\'aszl\'o Zalav\'ari}
\email{zalavarl@mcmaster.ca}
\affiliation{%
 Physics \& Astronomy, McMaster University, Hamilton, ON, Canada, L8S 4M1\\
 Perimeter Institute for Theoretical Physics, Waterloo, Ontario N2L 2Y5, Canada
}%

\date{\today}

\begin{abstract}
We apply point-particle effective field theory (PPEFT) to electronic and muonic $^4 \rm{He}^+$ ions, and use it to identify linear combinations of spectroscopic measurements for which the theoretical uncertainties are much smaller than for any particular energy levels. The error is reduced because these combinations are independent of all short-range physics effects up to a given order in the expansion in the small parameters $R/a_\ssB$ and $Z\alpha$ (where $R$ and $a_\ssB$ are the ion's nuclear and Bohr radii). In particular, the theory error is not limited by the precision with which nuclear matrix elements can be computed, or compromised by the existence of any novel short-range interactions, should these exist. These combinations of ${}^4$He measurements therefore provide particularly precise tests of QED. The restriction to ${}^4$He arises because our analysis assumes a spherically symmetric nucleus, but the argument used is more general and extendable to both nuclei with spin, and to higher orders in $R/a_\ssB$ and $Z\alpha$.
\end{abstract}

\maketitle 

\section{Introduction}
\label{sec:intro}

Atomic systems have historically been an important testing ground for Quantum Electrodynamics (QED), even providing one of the very first observations of a relativistic quantum effect with the Lamb Shift \cite{ref:LambShift}. Muonic atoms have further proved an excellent means of honing our understanding of QED by contrasting with electronic measurements. For muonic atoms, the leading QED radiative correction is due to electron-loop vacuum polarization \cite{Structure} in contrast to the electron's leading self-energy correction, and finite-size effects are enhanced by a factor $(m_\mu/m_e)^3 \sim 8\times 10^6$. Indeed, experiments in the 1970s found a discrepancy between theoretical and measured values for certain transitions in heavy muonic atoms \cite{ref:exp:Dixit1,ref:exp:Walter}. This motivated much research, and after a few years improvements in the theory \cite{ref:th:Sundaresan,ref:th:Blomqvist,ref:th:WiletsRinker,ref:th:RinkerWilets} and in experiments \cite{ref:exp:Kaeser,ref:exp:Dixit2} resolved the discrepancy and improved our understanding of QED \cite{ref:resolved}. Today a very similar situation can be found in the `proton-radius' problem \cite{ProtonRadius}, wherein the root-mean-squared charge radius inferred from the leading nuclear contributions to atomic energy shifts in Hydrogen and muonic Hydrogen appears to depend on the flavour of the orbiting lepton.

Recent laser spectroscopy of muonic atoms \cite{NewExp} has opened the door to new high-precision tests of QED, constituting tests of the theory at the two- and three-loop levels \cite{QEDShifts}. However, the small size of these higher-order QED corrections to atomic levels makes them compete with more mundane energy shifts, such as those due to the finite size of the nucleus. Consequently, uncertainties in computing nuclear contributions to atomic energy shifts are important components of the theoretical error budget when comparing with experiments. These theoretical uncertainties are made even worse if there should also be new short-range interactions between the nucleus and muon, such as have been motivated \cite{PRNew} by the proton-radius problem. Until it is understood whether this problem is solved by a better understanding of the experimental errors or through the existence of new physics, this discrepancy must be treated as an unknown unknown when assessing the theory error.

A better understanding of the nature of short-distance nucleus/lepton interactions is therefore an important prerequisite for exploiting the precision of spectroscopic measurements, both for the extraction of the best value of the Rydberg and to test QED. This is where effective field theory (EFT) in general \cite{QEDEFT, ProtonRadiusEFT}, and the point-particle effective field theory (PPEFT) framework in particular, can help \cite{PPEFT, KG, Dirac}. EFTs allow one to write a small set of effective interactions that capture the effects of {\em all} short-distance contributions to atomic energy levels (including both nuclear-scale physics and any hypothetical new short-range forces), order-by-order in powers of the relevant small size, $R$, of the physics in question. For nuclear physics $R$ would be of order the nuclear radius, while for a new short-range force it would instead be the force's range. The existence of these effective interactions allows a robust parameterization of the contributions of short-distance physics to atomic energy levels, without having to understand the details of its microscopic origin. 

Of course, knowing the underlying microscopic physics in question (such as the structure of the relevant nucleus) it becomes possible to compute the size of these effective interactions from first principles. In this language the uncertainties in nuclear-structure calculations enter into predictions through any inaccuracy in the values so inferred for the effective interactions. One of the points of this paper is to show how to relate such calculations to the effective couplings of the PPEFT framework in particular. 

We also take an entirely different tack. Instead of trying to reduce the inaccuracy of these effective couplings through more precise nuclear calculations, in this paper we also use the generality of the EFT parameterization to identify combinations of spectroscopic measurements from which all of the relevant short-distance effective couplings drop out to a fixed order in the expansion in $R/a_\ssB = m\epsilon Z\alpha$ and $Z\alpha$ (where $m$ is the mass of the orbiting particle, $Z$ is the nuclear charge, $\alpha$ is the fine-structure constant and $a_\ssB$ is the relevant Bohr radius). These combinations are particularly interesting because the absence of short-distance contributions to them means that the theoretical error for these observables is controlled by powers of $R/a_\ssB$ or $Z\alpha$ rather than by the larger uncertainties arising from (say) nuclear physics. A similar approach has been used to cancel dependence on nuclear effects for the hyperfine splitting in hydrogen \cite{hfs} (as well as to highlight nuclear isotope dependence, among other reasons \cite{QEDShifts, Lamberg}), however our approach has the advantage of being systematic, and can be applied in principle to any spinning or spinless nucleus. We can also extend our results to higher orders, as we illustrate by identifying nuclear-free combinations to higher order in $Z\alpha$ than has been done previously.

The key observation of this work is that the short-distance PPEFT couplings only enter into spectroscopic measurements through a single (mass-dependent) length-scale, $\epsilon_{\star e,\mu}$ (where the $e$ and $\mu$ are used to distinguish between the scale that applies to electrons versus muons). As a result, a single spectroscopic measurement for each fermion-type suffices to predict the finite-size contribution to all other energy shifts. Working to order $m^4R^3(Z\alpha)^5 \approx 10^{-2}\,\text{eV} \approx 10^3\, \text{GHz}$ for muonic atoms (as is relevant for the newest generation of muonic helium experiments\cite{NewExp}), we use this approach to predict
\begin{equation}
	\widehat{\Delta E}_{nS_{1/2}-nP_{1/2}} = \frac{8[2 + \alpha(\eta^{(\mu)}_{n0}-\eta^{(\mu)}_{n1})]}{n^3[2 + \alpha(\eta^{(\mu)}_{20}-\eta^{(\mu)}_{21})]}\; \widehat{\Delta E}_{2S_{1/2}-2P_{1/2}}\,, 
\end{equation}
and
\begin{align}
		\widehat{\Delta E}_{n_1S_{1/2}-n_2S_{1/2}} &= 2\widehat{\Delta E}_{2S_{1/2}-2P_{1/2}}\\
												   &\quad\times{}\left(\frac{2 + \alpha\eta^{(\mu)}_{n_1 0}}{n_1^2} - \frac{2 + \alpha\eta^{(\mu)}_{n_2 0}}{n_2^2}\right),\notag 
\end{align}
where
\be
  \widehat{\Delta E}_{1\to2} := \Delta E_{1\to2} - \Delta E^{\text{EM}}_{1 \to 2} = \Delta E^{\text{PP}}_{1 \to 2} + \Delta E^{\text{PP QED}}_{1 \to 2} 
\ee
is the difference between the total $n_1 X_{j1}$--$n_2 Y_{j2}$ transition ($\Delta E_{1\to 2}$) and the purely point-like contributions to the same difference ($\Delta E_{1\to 2}^{\text{EM}}$). (Equivalently, this is the difference between the finite-size correction to the $n_1 X_{j1}$ and $n_2 Y_{j2}$ states ($\Delta E^{\text{PP}}_{1 \to 2}$), plus the difference between the combined finite-size-QED corrections to the same states ($\Delta E^{\text{PP QED}}_{1 \to 2}$)). Here, $\eta_{n\ell}$ are computable $n$- and $\ell$-dependent coefficients associated with the combined finite-size-QED contributions given explicitly for electrons in \eqref{eq:EtaE} below.

For electrons, we also work to order $m^4R^3(Z\alpha)^5$, but now this is closer to $10^{-12}\,\text{eV} \approx 10^0\,\text{kHz}$, and so we must also include terms of order $m^3 R^2 (Z\alpha)^6 \approx 10^{-11}\,\text{eV}\approx 10\,\text{kHz}$, since the smaller electron mass makes those scales comparable. In this case, we predict for the same transitions:
\begin{align}
\widehat{\Delta E}_{nS_{1/2}-nP_{1/2}} &= \frac{8}{n^3}\widehat{\Delta E}_{2s_{1/2}-2p_{1/2}} \\
										&\quad\times {} \left\{1 + (Z\alpha)^2 \left[N(n) - \frac{n^2 - 1}{4n^2}\right] \right\} \,, \notag
\end{align}
and
\begin{align}
	\widehat{\Delta E}_{n_1 S - n_2 S} &= \widehat{\Delta E}_{2S_{1/2}-2P_{1/2}}\\
									   &\quad\times{}\left\{ \frac{1}{n_1^3} - \frac{1}{n_2^3} + (Z\alpha)^2\left[\frac{N(n_1)}{n_1^3} - \frac{N(n_2)}{n_2^3} \right]\right\},\notag
\end{align}
in which we define
\begin{equation}
	N(n) := \frac{12n^2-n-9}{4n^2(n+1)}  - H_{n+1} + \frac 54 + \frac{\eta^{(e)}_{n0}}{2Z} - \frac{\eta^{(e)}_{20}}{2Z} - \ln(\frac{2}{n})\,. 
\end{equation}

Moreover, even without solving for $\epsilon_\star$ explicitly, our knowledge of how this one parameter enters into energy shifts allows us to write down linear combinations of measurements from which it cancels altogether, thus defining relations between energy shifts that are entirely free of nuclear physics. For muons, we identify:
\begin{equation}
	\begin{split}
		&\frac{n_1^2}{2 + \alpha(\eta^{(\mu)}_{n_1 0} - \eta^{(\mu)}_{n_2 1})}\widehat{\Delta E}_{n_1S_{1/2} - n_1P_{3/2}} = \\
		&\qquad\qquad\frac{n_2^2}{2 + \alpha(\eta^{(\mu)}_{n_2 0} - \eta^{(\mu)}_{n_2 1})}\widehat{\Delta E}_{n_2S_{1/2} - n_2P_{3/2}} \,,
	\end{split}
\end{equation}
while for electrons, 
\begin{equation} 
	\begin{split}
		&\frac{24n_1^5}{n_1^2 - 1}\Delta E_{n_1P_{1/2}-n_1P_{3/2}} = \\
		&\qquad\frac{1}{F[n_1]-F[n_2]}\left(n_1^3\Delta E_{n_1S_{1/2}-n_1P_{1/2}} - n_2^3 \Delta E_{n_2S_{1/2}-n_2P_{1/2}}\right) 
	\end{split}
\end{equation}
where
\begin{equation}
	F[n]:= \frac{12n^2 - n - 9}{2n^2(n+1)} - \frac{n^2 - 1}{24n^2} + 2\ln n - 2H_{n+1} + \frac{\eta^{(e)}_{n0}}{Z}. 
\end{equation}

We organize our presentation as follows. \S\ref{sec:PPEFT} sets up the PPEFT framework required to draw the above conclusions, starting with a summary of the relevant near-nucleon boundary conditions and how these are related to the PPEFT effective description of the nucleus. This section also deals with various conceptual issues (such as deriving the appropriate renormalization-group (RG) invariant nuclear length scale $\epsilon_\star$. Next, \S\ref{sec:Comps} computes how this RG-invariant parameter captures various microscopic models for nuclei, including the moments of fixed charge distributions and nuclear polarizabilities. Once it is established how these contribute to atomic energy levels only through the one RG-invariant combination $\epsilon_\star$, we identify combinations of atomic transition frequencies from which this one nucleus-sensitive parameter cancels out. There are a great many such combinations, and each represents a quantity for which nuclear uncertainties are negligible at the level of present-day experimental measurements. \S\ref{sec:Nums} applies the formulae of the previous sections to the helium ion using the only available experimental data, the $2S_{1/2}-2P_{1/2}$ transition. The result is a prediction for the $1S-2S$ transition of $\nu_{1S-2S} = 9.868561009\, (1) \times 10^{9} \, \text{MHz}$, which is roughly 4 times less precise than predictions in the literature\cite{Yerokhin}. Our precision is however entirely dominated by the experimental error, and so can only improve with future experiments, while never relying on the inherently uncertain choice of a particular model of the nucleus. Finally, some conclusions are briefly summarized in \S\ref{sec:conc}.

\section{PPEFT for Spinless Nuclei}
\label{sec:PPEFT}

We present in this section a brief summary of EFT methods, as needed to discuss nuclear effects on the energy levels of electrons and muons orbiting spinless nuclei, such as ${}^4$He. EFTs are designed to exploit any hierarchies of scales in a problem to most efficiently compute a system's properties. As applied to atoms, EFTs such as NRQED \cite{QEDEFT} are usually used to exploit the hierarchy between the electron/muon mass and the much smaller size of typical bound-state energies. For PPEFT the hierarchy exploited is the large ratio between the small size, $R$, of the nucleus and the much larger size, $a_\ssB$, of the atomic Bohr radius. The expansion of observables in powers of $R/a_\ssB$ reveals them not to depend on most of the nuclear details, but only on a set of `generalized multipole moments', similar to the way that ordinary multipole moments control the expansion of the electrostatic field of a compact charge distribution. 

\subsection{PPEFT including Subleading Order}

This section reviews how to set up and solve for atomic energies within the PPEFT framework.

\subsubsection{Bulk System}

Before describing the nuclear degrees of freedom, we start by defining the long-distance, `bulk', fields whose properties the nucleus perturbs. We take the bulk system to be defined by QED, describing the renormalizable coupling of charge $-e$ fermions to photons,\footnote{Our metric has $(-+++)$ signature, so $\gamma^0$ is anti-hermitian while the spatial $\gamma^i$ are hermitian.}
\be
 S_\ssB = - \int \exd^4x \; \left[ \Psibar (\Dsl + m) \Psi + \frac14 \, F_{\mu\nu} F^{\mu\nu} \right] \,,
\ee
where $\Dsl = \gamma^\mu D_\mu$ with $\gamma^\mu$ denoting the usual Dirac gamma matrices and $D_\mu\Psi = (\partial_\mu + ieA_\mu ) \Psi$, as appropriate for fermions of charge $-e$, while $F_{\mu\nu} = \partial_\mu A_\nu - \partial_\nu A_\mu$. 

It is often useful to zoom in on the non-relativistic limit of this bulk physics by taking $m$ to be much larger than the energies of interest, and NRQED is the natural field-theoretic language for doing so. For later purposes it suffices to notice that this limit can be formally derived by performing a field redefinition that simplifies the large-$m$ limit. This is done for electrons and muons by redefining $\Psi \to \exp[mt\gamma^0  ]\Xi$, and assuming $\Xi$ to vary appreciably only over distances and times much larger than $1/m$. The point of this redefinition is to ensure $S_\ssB$ has a well-defined large-$m$ limit, since the term $m \Psibar \Psi = m\, \overline{\Xi} \,\Xi$ then precisely cancels the rest-mass part of the time derivative, $\Psibar \gamma^0 \partial_t \Psi = - m \,\overline{\Xi}\, \Xi + \cdots$, leaving interactions that can be expanded in powers of derivatives divided by $m$.

\subsubsection{Nuclear Properties}

If proceeding in the spirit of NRQED, nuclear properties could be included into the theory by adding its field, $\Phi$, preferably already within a non-relativistic framework that exploits expansions in inverse powers of the nuclear mass, $M$.

Within PPEFT, however, nuclear properties are instead identified by writing the {\it first-quantized} action for the nucleus that includes all possible local interactions between its centre-of-mass coordinate, $y^\mu(\tau)$, and the `bulk' fields $A_\mu(x)$ and $\Psi(x)$, respectively describing the electromagnetic potential and the Dirac field of the orbiting particle. This first-quantized framework is completely equivalent to the second-quantized one restricted to single-particle states and is more convenient when working purely within the single-nucleus sector, such as when describing an atom, for which most of the bells and whistles of quantum field theory for the nucleus are overkill.

For a spherically symmetric nucleus such as helium (or other doubly magic nuclei) restricted to parity-preserving interactions, this leads to \cite{Dirac}:
\begin{eqnarray} \label{eq:sourceaction0}
		S_p &=& -\int_\mathcal{W} \exd \tau \; \Bigl[ M - Ze \, A_\mu \dot y^\mu \nn\\
		&& \quad \qquad {} +  c_s \, \Psibar \,\Psi +i c_v \, \Psibar \,\gamma_\mu \Psi \, \dot y^\mu + \tilde h \, \dot y^\mu \partial^\nu F_{\mu\nu} \\
		&& \quad \qquad {} + id_s \, \dot y^\mu \Psibar \, D_\mu\Psi + d_v \, \dot y^\mu \dot y^\nu \Psibar \, \gamma_\mu D_\nu\Psi \nn\\
		&&\qquad {} +\frac12 (d_\ssE + d_\ssB) \dot y^\mu \dot y^\nu F_{\mu\lambda} {F_\nu}^\lambda + \frac 14 d_\ssB F_{\mu\nu}F^{\mu\nu} + \cdots \Bigr] \,.\nn
\end{eqnarray} 
Here $\mathcal{W}$ denotes the world-line $y^\mu(\tau)$ of the nuclear centre-of-mass --- along which $\tau$ is its proper time with derivative $\dot y^\mu := \exd y^\mu/\exd \tau$ --- at which all bulk fields are evaluated; as above $D_\mu\Psi = (\partial_\mu + ieA_\mu ) \Psi$.

The first line describes the physics of a point source with mass $M$ and charge $Ze$. The couplings $c_s$, $c_v$ and $\tilde h$ in the second line have dimensions of [length]$^2$, and so are expected to be order $R^2$ in size, up to dimensionless $\mathcal{O}(1)$ coefficients. Similarly the couplings $d_s$, $d_v$, $d_\ssE$ and $d_\ssB$ have dimension [length]${}^3$ and should be order $R^3$ and so on, with the ellipses containing all terms suppressed by more than three powers of $R$.

Since our focus is on energy shifts due to finite nuclear size, for simplicity of presentation we neglect kinematic nuclear recoil effects since the suppression of these corrections by powers of $m/M$ make their contributions to nuclear size effects smaller than the order to which we work. This amounts to assuming the nucleus to be at rest within the atomic rest frame: $\dot y^\mu = \delta^\mu_0$. Recoil corrections are, however, easily included within this framework by instead using (and quantizing) the full nuclear 4-velocity $\dot y^\mu = \gamma \{1, \bfv \}$, where $\gamma = (1 - \bfv^2)^{-1/2}$. 

With a static nucleus the above action becomes
\begin{eqnarray} \label{eq:sourceaction}
		S_p &=& -\int_\mathcal{W} \exd t \; \Bigl[ M - Ze \, A_0 \nn\\
		&& \qquad +  c_s \, \Psibar \,\Psi +i c_v \, \Psibar \,\gamma_0 \Psi  - \tilde h \, \nabla \cdot \bfE \\
		&&\quad + id_s \, \Psibar \, D_0 \Psi + d_v \, \Psibar \, \gamma_0 D_0 \Psi +\frac12 d_\ssE  \bfE^2 + \frac12 d_\ssB  \bfB^2 + \cdots \Bigr] \,.\nn
\end{eqnarray} 
In the absence of any $\Psi$ terms the four pure electromagnetic interactions would establish the particle to have electric charge $Ze$, charge-radius $r_p$ with $\tilde h = \frac16 \, Z e \, r_p^2$, and so on. The complete response of the atom to the nucleus, including nuclear polarizabilities \cite{Ji}, also requires direct couplings to $\Psi$, however, we see below how to relate these couplings to other nuclear properties, such as the polarizabilities and order-$R^3$ Friar moment contributions to the nuclear electrostatic form-factor \cite{Friar}. 

Because our interest is in largely non-relativistic applications for which kinematic effects arise as powers of $1/m$, just as for the bulk it can be convenient to rescale $\Psi = \exp[mt \gamma^0 ]\Xi$, to remove the rapidly-oscillating phase associated with the rest-mass. Having a reasonable large-$m$ limit after doing so requires the coefficients $c_s$ and $c_v$ to contain contributions proportional to $m$ that cancel those terms in $S_p$ involving time derivatives $\partial_t \Psi = m \gamma^0 \Psi + \cdots$, leading to  
\begin{equation}
	\begin{split}
		&c_s \, \Psibar \,\Psi +i c_v \, \Psibar \,\gamma_0 \Psi  + id_s \, \Psibar \, D_0 \Psi + d_v \, \Psibar \, \gamma_0 D_0 \Psi  \\
		&\qquad= (c_s - d_v m) \, \overline\Xi\,\Xi +i (c_v + d_s m) \, \overline\Xi \,\gamma_0 \Xi  +\cdots\,,
	\end{split}
\end{equation}
and so suggesting writing $c_s = d_v m + \tilde c_s$ and $c_v = - d_s m + \tilde c_v$, and so on. In what follows we make these replacements but drop the `tilde' on $c_s$ and $c_v$ to avoid notational clutter. Once this is done all time derivatives acting on $\Psi$ in $S_p$ can be treated as giving $\partial_t \Psi \to -i(\omega - m) \Psi$.

\subsubsection{Electromagnetic Response}

The purely electromagnetic terms in \eqref{eq:sourceaction} influence atomic energy levels through the change they introduce in the electromagnetic field sourced by the atomic nucleus. The naive way to compute the modified electric field represents the action \eqref{eq:sourceaction} as a delta-function, leading to the formal perturbative modification
\be \label{Efields}
  \bfE \simeq \bfE_c + \tilde h \, \nabla \delta^3(\bfr) + d_\ssE \bfE_c \, \delta^3(\bfr) \,,
\ee
in which $\bfE_c = (Ze/4\pi r^2) \hat \bfr$ denotes the lowest-order (Coulomb) field, with $\hat \bfr = \bfr/r$ being the radial unit vector.

What makes the above expression naive is the divergence of $\bfE_c$ at the support of the delta-function. A more precise way to formulate this (for which the PPEFT formalism is designed \cite{PPEFT, KG, Dirac}) is to recast the influence of $S_p$ on $A_\mu$ in terms of a boundary condition at a regularization surface at small but nonzero radius $r = \epsilon$. The couplings $\tilde h$ and $d_\ssE$ are regarded as depending implicitly on $\epsilon$ in such a way as to ensure that physical quantities do not depend on the precise value chosen for $\epsilon$.

What counts for energy shifts is the scalar potential implied by \eqref{Efields}. Keeping the regularization in mind the result is
\begin{equation}
	\label{eq:A0}
	A_0(r) = -\frac{Ze}{4\pi r} + \tilde h\, \delta^{(3)}(\bfr) + \frac{d_\ssE Ze}{(4\pi)^2\epsilon^2}\, f_\epsilon(r) \,,
\end{equation}
where the function $f_\epsilon(r)$ is any regularization consistent with $\nabla f_\epsilon = 4\pi \hat\bfr \, \delta^3 (\bfr)$ in the small-$\epsilon$ limit [such as $f_\epsilon(r) = - \Theta(\epsilon - r)/\epsilon^2$ where $\Theta(x)$ is the Heaviside step-function].

A similar story goes through for the magnetic field, for which the Maxwell equation gets modified by $S_p$ to become
\begin{equation} \label{eq:Max2_low}
	\nabla \times \mathbf{B} = d_\ssB\,\nabla\times[\mathbf{B}\,\delta^{(3)}(x))] \,.
\end{equation}
Because of the absence of nuclear spin (and so also magnetic moment) dictated by our spherical-symmetry assumption, nontrivial solutions to this arise only suppressed by powers of $v/c \sim Z \alpha$ and so are negligible to the order we work. This allows the neglect of the vector potential $\bfA$ in the calculations described below, in particular ensuring the magnetic polarizability, $d_\ssB$, contributes negligibly to atomic energies at the order we work.

To these must be added the corrections to the Dirac field due to the boundary-condition change it also experiences.

\subsubsection{Fermion Response}

To study atomic helium in this framework we examine QED involving the Dirac and electromagnetic quantum fields, subject to the boundary conditions implied \cite{PPEFT, KG, Dirac} by the presence of $S_p$. In this language it is only through these boundary conditions --- and the modification \eqref{eq:A0} --- that the nucleus affects atomic energy levels. More and more detailed nuclear contributions correspond to adding more and more complicated interactions to $S_p$, in what amounts to a `generalized multipole expansion' of the nucleus.

In this framework QED interactions are included perturbatively as usual, with bound-state energies obtained from the positions of poles of the two-point function $\langle 0 | T \Psi(x) \Psibar(x') | 0 \rangle$. These are determined in part by computing the modes $\psi_n(x) = \langle 0 | \Psi(x) | n \rangle$ everywhere away from the nucleus. Perturbation theory is set up as usual, with the unperturbed system neglecting QED and nuclear corrections to $A_0$: {\it i.e.}~using for $\psi_n$ solutions to the Dirac equation with a Coulomb potential:  
\begin{equation}
	(\slsh D + m)\psi = \left[ -\gamma^0\left( \omega + \frac{Z\alpha}{r}  \right) + \vec{\gamma}\cdot\vec\grad + m\right]\psi = 0 \,,
\end{equation}
for energy eigenstates $\psi \propto e^{-i\omega t}$. 

This has well-known solutions of definite parity and total angular momentum given by:
\begin{equation}\label{eq:bulk}
\psi^\pm = \left( \begin{array}{c}  f_\pm(r) \,U^\pm_{j j_z}(\theta,\phi) +i g_\pm(r) \,U^\mp_{j j_z}(\theta,\phi) \\  f_\pm(r) \,U^\pm_{j j_z}(\theta,\phi) -i g_\pm(r) \,U^\mp_{j j_z}(\theta,\phi) \end{array}  \right)  \,,
\end{equation}
where $\psi^+$ and $\psi^-$ denote parity eigenstates, $U^\pm_{j j_z}$ are the Dirac spinor harmonics with definite total angular momentum $j = \ell \pm \frac 12$ and the parity eigenvalue is $\hat\Pi\, U^\pm_{j j_z} = (-)^{j \mp \frac 12}U^\pm_{j j_z}$.

To lowest order the functions $f_\pm(r)$ and $g_\pm(r)$ solve the radial part of the Dirac-Coulomb equation, and for a source with charge $Ze$ have the form
\begin{widetext}
\bea \label{eq:fACgentxt}
    f_\pm &=& \sqrt{m+\omega}\,  e^{-\rho/2} \rho^{\zeta-1} \left\{ A_\pm \, \cM \left[ \zeta -\frac{Z\alpha \omega}{\kappa}, 2\zeta+1; \rho \right] +  C_\pm \rho^{-2\zeta} \cM \left[ -\zeta -\frac{Z\alpha \omega}{\kappa}, -2\zeta+1; \rho \right] \right. \\
		  &&{} \left. -  A_\pm \left(\frac{\zeta-{Z\alpha \omega}/{\kappa}}{K-{Z\alpha m}/{\kappa}}\right) \cM \left[ \zeta -\frac{Z\alpha \omega}{\kappa}+1, 2\zeta+1; \rho \right] + C_\pm \left(\frac{\zeta+{Z\alpha \omega}/{\kappa}}{K-{Z\alpha m}/{\kappa}}\right) \rho^{-2\zeta} \cM \left[ -\zeta -\frac{Z\alpha \omega}{\kappa}+1, -2\zeta+1; \rho \right] \right\} \,, \nn
\eea
and
\bea \label{eq:gACgentxt}
    g_\pm &=& -\sqrt{m-\omega}\,  e^{-\rho/2} \rho^{\zeta-1} \left\{ A_\pm \, \cM \left[ \zeta -\frac{Z\alpha \omega}{\kappa},2\zeta+1;\rho \right] + C_\pm \rho^{-2\zeta} \cM \left[ -\zeta -\frac{Z\alpha \omega}{\kappa},-2\zeta+1;\rho\right] \right. \\
		  &&{} \left. + A_\pm \left( \frac{\zeta-{Z\alpha \omega}/{\kappa}}{K-{Z\alpha m}/{\kappa}} \right) \cM \left[\zeta -\frac{Z\alpha \omega}{\kappa}+1,2\zeta+1;\rho\right] - C_\pm \left( \frac{\zeta+{Z\alpha \omega}/{\kappa}}{K-{Z\alpha m}/{\kappa}}\right) \rho^{-2\zeta} \cM \left[-\zeta -\frac{Z\alpha \omega}{\kappa}+1,-2\zeta+1;\rho\right] \right\} \,,\nn
\eea
\end{widetext}
where $A_\pm$ and $C_\pm$ are integration constants, $\cM[a,b; z] = 1 + (a/b)z + \cdots$ are the standard confluent hypergeometric functions, $\omega$ is the mode energy while $\rho = 2 \kappa r$ where $\kappa$ and $\zeta$ are defined by
\be
  \kappa = \sqrt{(m-\omega)(m+\omega)}  \quad \hbox{and} \quad \zeta = \sqrt{\left(j+\frac12\right)^2 - (Z\alpha)^2} \,.
\ee
In what follows, $\kappa$ is real because we study atomic bound states which satisfy $m > \omega$. The parity of the state often enters through the parameter $K= \mp(j+\frac12)$ where the upper (lower) sign in $K$ corresponds to state $\psi^+$ (or $\psi^-$).

In this language the entire influence of nuclear-scale physics on the orbiting fermion arises through the boundary condition implied by the point-particle action \eqref{eq:sourceaction} for the bulk fields $\Psi$ and $A_\mu$ near the origin \cite{PPEFT, KG, Dirac}. Nuclear contributions to QED corrections similarly enter through the boundary conditions satisfied by the propagators built from these modes in the relevant graphs. 

\subsubsection{Near-nucleus boundary conditions}

The main result (explained in some detail for the Dirac equation in \cite{Dirac}) governing how nuclear properties perturb atomic levels relates the parameters of $S_p$ to the near-nucleus value of the ratios $(g_+/f_+)_{r=\epsilon}$ and $(f_-/g_-)_{r=\epsilon}$ of the radial modes evaluated at a small (but arbitrary) distance $r = \epsilon$ outside the nucleus: $R \le \epsilon \ll a_\ssB$ (with $R$ the smallest radius where an external extrapolation is valid and $a_\ssB$ denoting the relevant atomic Bohr radius). The ratios $g_+/f_+$ and $f_-/g_-$ at $r = \epsilon$ determine the physical integration constant that arises in the general solution to the radial equation, which in turn controls the dependence of atomic observables.\footnote{Notice that specifying $f_\pm/g_\pm$ at $r=\epsilon$ generically implies the radial functions need not remain bounded at the origin, which is the traditional choice for boundary conditions there. But this is not a fundamental worry because the growth of the radial solution eventually gets cut off once the interior of the nucleus is reached and the asymptotic solution of the Coulomb Dirac equation no longer approximates the real physics.}

It is convenient when stating the boundary conditions to write the ratios $(g_\pm/f_\pm)_{r=\epsilon}$ in a way that makes manifest the small parameters in the problem: the two small quantities $\epsilon/a_\ssB = m\epsilon Z\alpha$ and $(Z\alpha)^2$. This is most conveniently done by writing:
\begin{equation}\label{eq:rats}
	\left( \frac{g_+}{f_+} \right)_{r = \epsilon} = \xi_g\, Z\alpha, \quad \text{and} \quad X\left( \frac{f_-}{g_-} \right)_{r = \epsilon} = \frac {\xi_f}{2n},
\end{equation}
where $X := \sqrt{(m-\omega)/(m+\omega)}$ is included for later notational simplicity, while $n$ is the state's principal quantum number and or atomic energy levels, $\omega = m - (Z\alpha)^2 m/(2n^2) + \cdots$. The quantities $\xi_f$ and $\xi_g$ then have the expansions 
\begin{align}\label{eq:hats}
		&\xi_g := \hat g_1(\epsilon) +  (m\epsilon Z\alpha) \hat g_2(\epsilon) + (Z\alpha)^2  \hat g_3(\epsilon) + \dots \,, \notag \\
		&\xi_f := (m\epsilon Z\alpha) \hat f_1(\epsilon)+  (m\epsilon Z\alpha)^2 \hat f_2(\epsilon)+ (Z\alpha)^2 \hat f_3(\epsilon)+ \dots \,,
\end{align}
where the ellipses involve terms involving more powers of $(m\epsilon Z\alpha)$ and/or $(Z\alpha)^2$ than those written, and the dependence on $n$ follows directly from the $\omega$-dependence of the radial Dirac equation. The dimensionless coefficients $\hat g_i(\epsilon)$ and $\hat f_i(\epsilon)$ are normalized in \eqref{eq:rats} so as to ensure that $\hat g_1$ are order-unity in applications to atomic energy levels. 

\subsubsection{Energy shifts}

Before determining how $\hat f_i$ and $\hat g_i$ depend on nuclear parameters, we briefly summarize how these quantities are related to shifts in atomic energy levels. 

As shown in detail in \cite{Dirac}, the ratio of integration constants $A_\pm/C_\pm$ appearing in the solutions \eqref{eq:fACgentxt} and \eqref{eq:gACgentxt} can be determined if $f_\pm/g_\pm$ is regarded as being specified at $r = \epsilon$. For bound states, imposing normalizability at infinity over-determines the eigenvalue problem in the usual way, leading to standard predictions for the bound-state energy levels. Writing the shift in these energies relative to the standard Dirac energies (obtained when $C_\pm = 0$) due to the deviations in $\hat f_i$ and $\hat g_i$ \cite{Dirac} as $\delta E$, gives the nucleus-dependent shift to the $j=\frac12$ positive- and negative-parity energy levels as:
\ba\label{eq:even}
		\delta E^+_{1/2} &\simeq& \frac{m^3 \epsilon^2 (Z\alpha)^4}{n^3} \Bigl\{ 2(1+2\hat g_1)   \\
						 &&{} + \left[ 2\hat g_2 - \frac{8}{3} - 4\,\hat g_1(\hat g_1+2)\right](m\epsilon Z\alpha) \notag \\
				&&{} + \left[ 4\hat g_3+5 +8\hat g_1  -2\hat g_1^2  + (1+2\hat g_1)\left\{\frac{12n^2-n-9}{2n^2(n+1)} \right.\right.\notag \\
				&&{}\left.\left.- 2\ln \left( \frac{2m\epsilon Z\alpha}{n} \right) - 2H_{n+1} - 2\gamma\right\} \right](Z\alpha)^2 +\dots\Bigr\}\,,\nn
 \ea
for parity-even states and
 \begin{equation}\label{eq:odd}
	 \begin{split}
		 \delta E^-_{1/2}  &\simeq  -\frac{(n^2-1)}{n^5} m^4 \epsilon^3 (Z\alpha )^5 \left( \hat f_1  - \frac23\right) \\
						&\quad{}+\frac{n^2 - 1}{2n^5}m^3 \epsilon^2(Z\alpha)^6\left(1 - 2\hat f_3\right) + \cdots\,,
	 \end{split}
 \end{equation}
for parity-odd states.  In these expressions the ellipses contain terms suppressed by higher powers of $(m\epsilon Z\alpha)$ and $(Z\alpha)^2$. Here $\gamma$ is the Euler-Mascheroni constant and $H_n$ are the harmonic numbers, $H_m = 1 + \frac 12 + \frac 13 + \dots + \frac 1 m$, and so $H_1 = 1$, $H_2 = \frac32$, $H_3 = \frac{11}6$ and so on. 

We include in the above all contributions relevant to the current generation of experiments involving electrons and muons orbiting a ${}^4$He nucleus. Recall that for muons, $(m\epsilon Z\alpha) \gg (Z\alpha)^2$ when $\epsilon$ is a typical nuclear size, but for electrons $(m\epsilon Z\alpha) \simeq (Z\alpha)^2$.  Consequently, for muonic atoms it suffices to keep terms of order $m^4 \epsilon^3(Z\alpha)^5$ while dropping terms of size $m^3 \epsilon^2 (Z\alpha)^6$, but for electrons these terms must both be kept. This means the coefficients $\hat g_1$, $\hat g_2$ and $\hat f_1$ are in principle of interest for muonic He, while all of $\hat g_1$, $\hat g_2$, $\hat g_3$, $\hat f_1$ and $\hat f_3$ are relevant for electrons. It is for this reason that the contribution to $\delta E^-$ from $\hat f_2$ is not written in \eqref{eq:odd}. Similarly, the leading contributions for $j = \frac32$ are the same size as terms neglected above, and so can be dropped in what follows.

Later sections evaluate these formulae using $\hat f_i$ and $\hat g_i$ as computed with several simple specific models of nuclei, and in this way we verify that they include the results of standard calculations in the literature. In particular, they contain the various moments encountered when doing so with the nucleus modelled as a static charge distribution, reducing to well-known formulae for finite-size corrections to the Dirac-Coulomb energies \cite{Dirac, Structure, Eides, Zemach, Friar, Nickel, NASA}. 
However, as we see below, the real power of the above expressions \eqref{eq:even} and \eqref{eq:odd} is in their generality since once computed in terms of the parameters in $S_p$ they capture the effects of arbitrary short-distance physics localized at the nucleus.\footnote{The interactions of $S_p$ specialize to rotational and parity invariance, but nothing in principle forbids extending these interactions to include nuclear spin and parity-violating interactions.}

\subsection{Matching and RG Invariance}

The influence of the nucleus on atomic levels (or on low-energy lepton scattering) is completely determined by the near-nucleus boundary condition for the modes $\psi$ at $r = \epsilon$, and so is ultimately parameterized by the dependence of the coefficients $\hat g_i(\epsilon)$ and $\hat f_i(\epsilon)$ on nuclear parameters. The mapping of nuclear physics to atomic physics is completely captured by describing this dependence, and the point of the PPEFT formalism is to parameterize this dependence efficiently so as to exploit the hierarchy of scales $R \lesssim \epsilon \ll a_\ssB$. 

\subsubsection{Connecting Boundary Conditions to \texorpdfstring{$S_p$}{S\_p}}

The main consequence of $S_p$ for atomic levels comes from the boundary condition it implies at $r = \epsilon$ for the radial functions $f_\pm(r)$ and $g_\pm(r)$. These are worked out at leading nontrivial order in \cite{Dirac}, and the result is extended to include the subdominant interactions of \eqref{eq:sourceaction} in \cite{Dirac2}. The boundary conditions that follow from these references are
\begin{equation}
\label{eq:psiBC1}
	\begin{split}
	&\hat c^\prime_s + \hat c^\prime_{v\, \text{tot}} -\frac{(Z\alpha)}{2n^2} (\hat d_s + \hat d_v)(m\epsilon Z\alpha) = \left( \frac {g_+}{f_+} \right)_{r=\epsilon} \\
&\qquad= Z\alpha \Bigl[ \hat g_1(\epsilon) + (m\epsilon Z\alpha) \hat g_2(\epsilon) +  (Z\alpha)^2 \hat g_3(\epsilon) + \dots \Bigr] \,,
	\end{split}
\end{equation}
for the parity-even states and
\begin{equation}
\label{eq:psiBC2}
	\begin{split}
		&\hat c^\prime_s - \hat c^\prime_{v\, \text{tot}} -\frac{(Z\alpha)}{2n^2} (\hat d_s - \hat d_v)(m\epsilon Z\alpha)    
   = \left( \frac {f_-}{g_-} \right)_{r=\epsilon}  \\
   &\qquad= \frac{1}{2nX} \left[(m\epsilon Z\alpha)  \hat f_1(\epsilon)+  (m\epsilon Z\alpha)^2 \hat f_2(\epsilon)\right. \\
   &\qquad\qquad\qquad\,\, \left.+  (Z\alpha)^2 \hat f_3(\epsilon)+ \dots \right] \,,
	\end{split}
\end{equation}
for the parity-odd states. Here the hatted quantities are $\hat c_{s, v\,\text{tot}}^\prime := c_{s, v\,\text{tot}}^\prime/4\pi\epsilon^2$ while $\hat d_{s,v} := d_{s,v}/4\pi\epsilon^3$, and so are dimensionless. Primes denote the combinations 
\begin{equation}
     c^\prime_{s,v} := c_{s,v} - \frac{(Z\alpha)d_{s,v}}{\epsilon} \,.
\end{equation}
Finally, the subscript `tot' represents the combination
\begin{equation}
	c_{v\,\text{tot}} := c_v - e\tilde h - \frac{d_\ssE Z\alpha}{3\epsilon} \,.
\end{equation}
The parameters $c_v$ and $e\tilde h$ naturally combine in this way, since both of these effective interactions introduce a delta-function potential in the non-relativistic Schr\"odinger limit \cite{PPEFT, Dirac}. 

The final step is to solve the above boundary condition to relate the quantities $\hat f_i$ and $\hat g_i$ to the parameters $c_s$, $c_v$, $d_s$, $d_v$, $\tilde h$ and $d_\ssE$. This allows a determination of which nuclear parameters govern which atomic energy shifts. Before doing so we first show how to deal with the apparent arbitrariness associated with the ubiquitous $\epsilon$-dependence of the boundary conditions. Doing so allows an efficient identification of the physical quantities, and in particular allows a clean counting of the number of nuclear parameters that enter into energy shifts at any given order.

\subsubsection{Renormalization-Group Running}

Recall that the position, $r = \epsilon$, where the boundary conditions \eqref{eq:psiBC1} and \eqref{eq:psiBC2} are imposed is basically arbitrary, so long as it lies outside the nucleus and is much smaller than the atomic Bohr radius.  This makes it odd that expressions like \eqref{eq:even} and \eqref{eq:odd} for physical energy shifts appear to make them depend on $\epsilon$.  The purpose of this section is to show why this dependence is really an illusion, because it is cancelled by an $\epsilon$-dependence that is implicit in the effective couplings $c_s$, $c_v$ and so on. This section develops renormalization-group (RG) tools for determining this dependence explicitly, thereby allowing a determination of the physical RG-invariant content of the effective couplings. 

To this end it is important to realize that equations like \eqref{eq:psiBC1} and \eqref{eq:psiBC2} can be read in two ways. First, it can be read as giving the $\epsilon$-dependence required of the effective couplings in order to ensure that physical quantities remain $\epsilon$-independent. This is done by equating it to the $\epsilon$-dependence that is explicit on the right-hand side (through the evaluation of the bulk solution for $f_\pm/g_\pm$). The condition that physical quantities be independent of $\epsilon$ in this language corresponds to demanding that the ratio of integration constants, $A_\pm/C_\pm$, be $\epsilon$-independent (and so RG invariant as $\epsilon$ is varied).   

Once this is done, the $\epsilon$-dependence on both sides of eqs.~\eqref{eq:psiBC1} and \eqref{eq:psiBC2} becomes identical, and then the second way to read these equations is to equate the RG-invariant coefficients on both sides. This then gives the ratio of integration constants $A_\pm/C_\pm$ in terms of RG-invariant parameters. But because energy shifts can be computed from $A_\pm/C_\pm$ this also gives predictions for the energy shifts in terms of the RG invariant characterizations of the coupling flow.

To start this off we take the small-$r$ asymptotic form of the solutions given in \eqref{eq:fACgentxt} and \eqref{eq:gACgentxt} and use these to evaluate $g_+/f_+$ and $f_-/g_-$ on the right-hand sides of eqs.~\eqref{eq:psiBC1} and \eqref{eq:psiBC2}. This leads to the following expressions:
\begin{align}
	\label{eq:newRunBase+}
	&\hat c^\prime_s + \hat c^\prime_{v\, \text{tot}} -\frac{(Z\alpha)}{2n^2} (\hat d_s + \hat d_v)(m\epsilon Z\alpha) = \\
	&\qquad\qquad-X \frac{\left\{ \left( g^{+}_{02} + g^+_{03}\rho \right) + \left( g^+_{12} + g^+_{13}\rho \right)\frac{C_+}{A_+}\rho^{-2\zeta} \right\}}{\left\{ \left( f^{+}_{02} + f^+_{03}\rho \right) + \left( f^+_{12} + f^+_{13}\rho \right)\frac{C_+}{A_+}\rho^{-2\zeta} \right\}}, \notag 
\end{align}
and
\begin{align}
	\label{eq:newRunBase-}
	&\hat c^\prime_s - \hat c^\prime_{v\, \text{tot}} -\frac{(Z\alpha)}{2n^2} (\hat d_s + \hat d_v)(m\epsilon Z\alpha) = \\
	&\qquad\qquad-\frac{1}{X} \frac{\left\{ \left( f^{-}_{02} + f^-_{03}\rho \right) + \left( f^-_{12} + f^-_{13}\rho \right)\frac{C_-}{A_-}\rho^{-2\zeta} \right\}}{\left\{ \left( g^{-}_{02} + g^-_{03}\rho \right) + \left( g^-_{12} + g^-_{13}\rho \right)\frac{C_-}{A_-}\rho^{-2\zeta} \right\}}, \notag
\end{align}
where (as before)  $X := \sqrt{(m - \omega)/(m + \omega)}$ while $\rho = 2\kappa \epsilon = 2m\epsilon\sqrt{1 - \omega^2/m^2} \simeq 2m\epsilon Z\alpha/n$. Finally, the coefficients are given by
\begin{widetext}
\begin{equation} \label{g0212etc}
	\begin{alignedat}{2}
		& g^+_{02} := -\left(j + \frac12\right) + \zeta - \frac{Z\alpha}{X} \qquad\qquad && g^+_{12} := -\left(j + \frac12\right) - \zeta - \frac{Z\alpha}{X} , \\
		& f^+_{02} := -\left(j + \frac12\right) - \zeta - Z\alpha X \qquad\qquad && f^+_{12} := -\left(j + \frac12\right) + \zeta - Z\alpha X,
	\end{alignedat}
\end{equation}
and
\begin{equation}\label{f0212etc}
	\begin{alignedat}{2}
		& f^-_{02} := \left(j+\frac12\right) - \zeta - Z\alpha X \qquad\qquad && f^-_{12} := \left(j+\frac12\right) + \zeta - Z\alpha X, \\
		& g^-_{02} := \left(j+\frac12\right) + \zeta - \frac{Z\alpha}{X} \qquad\qquad && g^-_{12} := \left(j+\frac12\right) - \zeta - \frac{Z\alpha}{X}.
	\end{alignedat}
\end{equation}
and
\begin{equation}
	\begin{alignedat}{2}
		& g^+_{03} := \frac{\left(\zeta - Z\alpha \omega/\kappa\right)\left(\zeta - Z\alpha/X\right)}{2\zeta + 1}\qquad\qquad && g^+_{13} := \frac{\left(\zeta + Z\alpha\omega/\kappa\right)\left(\zeta + Z\alpha/X\right)}{-2\zeta + 1}, \\
		& f^+_{03} := \frac{\left(\zeta - Z\alpha\omega/\kappa\right)\left(-\zeta - Z\alpha X - 2\right)}{2\zeta + 1} \qquad\qquad && f^+_{13} := \frac{\left(\zeta + Z\alpha\omega/\kappa\right)\left(-\zeta + Z\alpha X + 2\right)}{-2\zeta + 1}, \\
		& f^-_{03} := \frac{\left(\zeta - Z\alpha\omega/\kappa\right)\left(-\zeta - Z\alpha X\right)}{2\zeta + 1} \qquad\qquad && f^-_{13} := \frac{\left(\zeta + Z\alpha\omega/\kappa\right)\left(-\zeta + Z\alpha X \right)}{-2\zeta + 1}, \\
		& g^-_{03} := \frac{\left(\zeta - Z\alpha \omega/\kappa\right)\left(2 + \zeta - Z\alpha/X\right)}{2\zeta + 1} \qquad\qquad && g^-_{13} := \frac{\left(\zeta + Z\alpha\omega/\kappa\right)\left(-2 + \zeta + Z\alpha/X\right)}{-2\zeta + 1}.
	\end{alignedat}
\end{equation}
	
\end{widetext}

These equations show that it is the series in integer powers of $\rho$ on the right-hand-side that corresponds to the expansion in powers of $m\epsilon Z \alpha$ on the left-hand side. Temporarily working to lowest order in this expansion leads to the expression found in \cite{Dirac} for the running of the couplings $\hat c_s$ and $\hat c_{v,\,\text{tot}}$:
\be \label{eq:origRun+}
	\hat c_s + \hat c_{v,\,\text{tot}} = -X \left( \frac{g^+_{02} + g^+_{12}\frac{C_+}{A_+}\rho^{-2\zeta}}{f^+_{02} + f^+_{12}\frac{C_+}{A_+}\rho^{-2\zeta}} \right), 
\ee
with coefficients given in \eqref{g0212etc}. Similarly,
\begin{align}
	\label{eq:origRun-}
	\hat c_s - \hat c_{v,\,\text{tot}} &= -\frac{1}{X} \left( \frac{f^-_{02} + f^-_{12}\frac{C_-}{A_-}\rho^{-2\zeta}}{g^-_{02} + g^-_{12}\frac{C_-}{A_-}\rho^{-2\zeta}} \right),
\end{align}
with coefficients given in \eqref{f0212etc}. 

These expressions give the RG-evolution of $\hat c^\prime_s \pm \hat c^\prime_{v,\,\text{tot}}$ as functions of $\epsilon$. It is convenient to rewrite the result as
\begin{equation}
	\label{eq:origRunComb}
	\hat c^\prime_s \pm \hat c^\prime_{v,\,\text{tot}} = \bar \lambda_\pm,
\end{equation}
where the $\bar \lambda^\pm$ are given by
\begin{equation}
	\label{eq:defBarLam}
	\bar\lambda_\pm := \frac{1}{Z\alpha}\left[\pm\zeta\frac{(\epsilon/\epsilon^\pm_\star)^{2\zeta} + \eta_\pm}{(\epsilon/\epsilon^\pm_\star)^{2\zeta} - \eta_\pm} + K\right]\,,
\end{equation}
where $K := \mp (j + 1/2)$, with upper (lower) sign corresponding to parity even (odd). Eq.~\eqref{eq:defBarLam} defines two types of RG evolution, distinguished by the parameter $\eta_\pm := \text{sgn}(\abs{(Z\alpha)\bar\lambda_\pm -K} - 1)$. $\eta_\pm = 1$ corresponds to a class of evolution for which $\bar\lambda_\pm$ never passes through $-K/Z\alpha$ and is unbounded (diverging at $\epsilon = \epsilon^\pm_\star$). $\eta_\pm = -1$ represents a class of evolution for which $\bar\lambda_\pm$ is bounded and passes through $-K/Z\alpha$ once (at $\epsilon = \epsilon^\pm_\star$). 

This evolution can also be recast in differential form by differentiating while requiring $C_\pm/A_\pm$ to be $\epsilon$-independent, and re-expressing the result in terms of $\bar\lambda_\pm$. Eq.~\eqref{eq:defBarLam} trades the constants $C_\pm/A_\pm$ for convenient RG-invariant integration constants, $\epsilon_\star^\pm$, obtained by integrating the differential evolution.

How is this picture changed once we include the $m\epsilon Z \alpha$ corrections? It turns out that the functions $\bar\lambda_\pm(\epsilon)$ are very useful in this case too, because the functional form \eqref{eq:defBarLam} appears in the coefficients of each power of $\rho$ in \eqref{eq:newRunBase+} and \eqref{eq:newRunBase-}.  In particular, the generalization of \eqref{eq:origRun+} and \eqref{eq:origRun-} to next order in $m\epsilon Z \alpha$ has the form 
\begin{equation}
	\label{eq:fullRun}
	\begin{split}
		&\hat c^\prime_s \pm \hat c^\prime_{v\, \text{tot}} - \frac{(Z\alpha)}{2n^2} (\hat d_s \pm \hat d_v)(m\epsilon Z\alpha) = \\
		&\qquad\qquad\bar\lambda_\pm + \frac{1}{n}[C_0^\pm + C_1^\pm\bar\lambda_\pm + C_2^\pm \bar\lambda^2_\pm](2m\epsilon Z\alpha),
	\end{split}
\end{equation}
where, evaluating $X$ and $\kappa$ using the lowest-order Coulomb energy, $\omega/m \approx 1 - \frac12(Z\alpha/n)^2$, 
\begin{equation}
	\label{eq:defABC}
	\begin{split}
		C_0^+ &:= \frac{X(g^+_{02}g^+_{13}-g^+_{03}g^+_{12})}{f^+_{02}g^+_{12} - f^+_{12}g^+_{02}} \approx \frac{8n^2+1}{12n}(Z\alpha) + \dots,  \\
		C_1^+ &:= \frac{f^+_{02}g^+_{13} - f^+_{03}g^+_{12} - f^+_{12}g^+_{03} + f^+_{13}g^+_{02}}{f^+_{02}g^+_{12} - f^+_{12}g^+_{02}} \approx 2n + \dots, \\
		C_2^+ &:= \frac{f^+_{02}f^+_{13} - f^+_{03}f^+_{12}}{X(f^+_{02}g^+_{12} - f^+_{12}g^+_{02})} \approx \frac{n}{Z\alpha} + \frac{8n^2 - 2n + 1}{4n}(Z\alpha)+\dots\,,
	\end{split}
\end{equation}
and
\begin{equation}
	\label{eq:defABC-}
	\begin{split}
		C_0^- &:= \frac{f^-_{02}f^-_{13}-f^-_{03}f^-_{12}}{X(g^-_{02}f^-_{12} - g^-_{12}f^-_{02})} \approx \frac{n}{3(Z\alpha)} + \dots, \\
		C_1^- &:= \frac{g^-_{02}f^-_{13} - g^-_{03}f^-_{12} - g^-_{12}f^-_{03} + g^-_{13}f^-_{02}}{g^-_{02}f^-_{12} - g^-_{12}f^-_{02}} \approx \frac{2n}{3} + \dots, \\
		C_2^- &:= \frac{X(g^-_{02}g^-_{13} - g^-_{03}g^-_{12})}{g^-_{02}f^-_{12} - g^-_{12}f^-_{02}} \approx -\frac{8n^2 - 3}{12n}(Z\alpha)+\dots \,. 
	\end{split}
\end{equation}
Here ellipses indicate higher powers of $Z\alpha$. 

Equating the coefficients of each power of $m\epsilon Z \alpha$ in \eqref{eq:fullRun} dictates separately the running of $\hat c^\prime_s$, $\hat c^\prime_{v,\,\text{tot}}$ (given by \eqref{eq:origRunComb}), and $\hat d_s$ and $\hat d_v$. The running of $\hat d_s$ and $\hat d_v$ is given by:
\begin{align}
	\label{eq:runD+}
	Z\alpha(\hat d_s + \hat d_v) &= -\frac{8n^2 + 1}{3}(Z\alpha) - 8n^2\bar\lambda_+ \\
								 &\qquad{}-\left(\frac{4n^2}{Z\alpha} + (8n^2 - 2n + 1)(Z\alpha)\right) \bar\lambda_+^{2}, \notag
\end{align}
and
\begin{align}
	\label{eq:runD-}
	Z\alpha(\hat d_s - \hat d_v)  &= -\frac{4n^2}{3(Z\alpha)} - \frac{8n^2}{3}\bar\lambda_- + \frac{8n^2 - 3}{3}(Z\alpha)\bar\lambda_-^{2}\,.
\end{align}
Interestingly, the running of all of the effective couplings are controlled by the two functions $\bar\lambda_\pm(\epsilon)$. As a result the flow of all couplings is described in principle by the same two RG-invariant constants, $\epsilon_\star^\pm$. These two parameters encode the information contained in $C_\pm/A_\pm$ in the solutions $f_\pm$ and $g_\pm$. As we see below, only one of these two quantities is independent for a parity-preserving nucleus since $\epsilon^+_\star = \epsilon^-_\star =: \epsilon_\star$. 

\begin{figure}
\begin{center}
\includegraphics[width=0.47\textwidth]{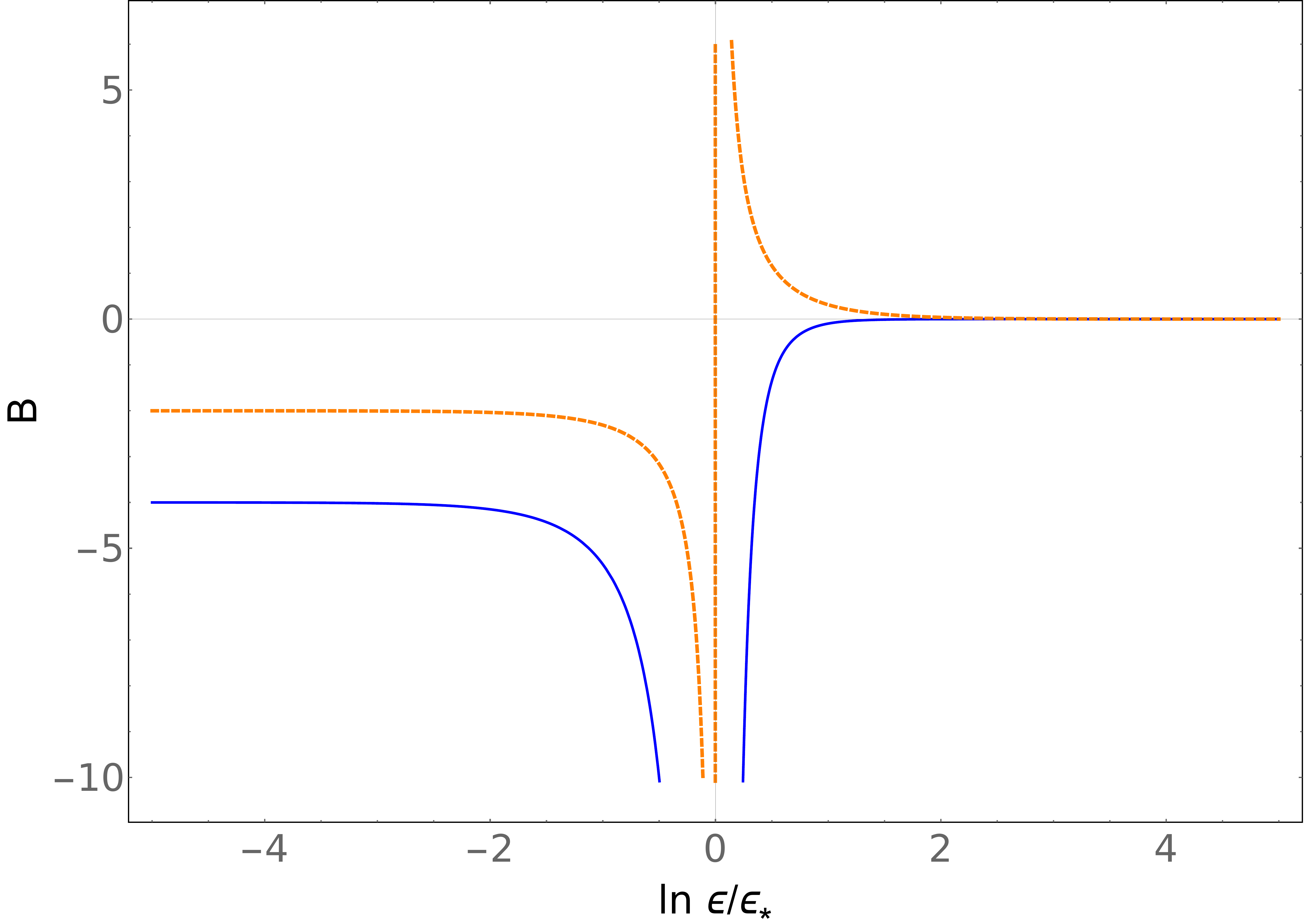} 
\caption{Plot of the RG flow of $B = (Z\alpha)^4(\hat d_s + \hat d_v)/4$ [solid blue] and $B = (Z\alpha)(\hat c^\prime_s + \hat c^\prime_{v\,\text{tot}})$ [dashed orange] vs $\ln \epsilon/\epsilon_\star$, with $\eta_+ = +1$.}\label{fig:class1} 
\end{center}
\end{figure}

\begin{figure}[ht]
\begin{center}
\includegraphics[width=0.47\textwidth]{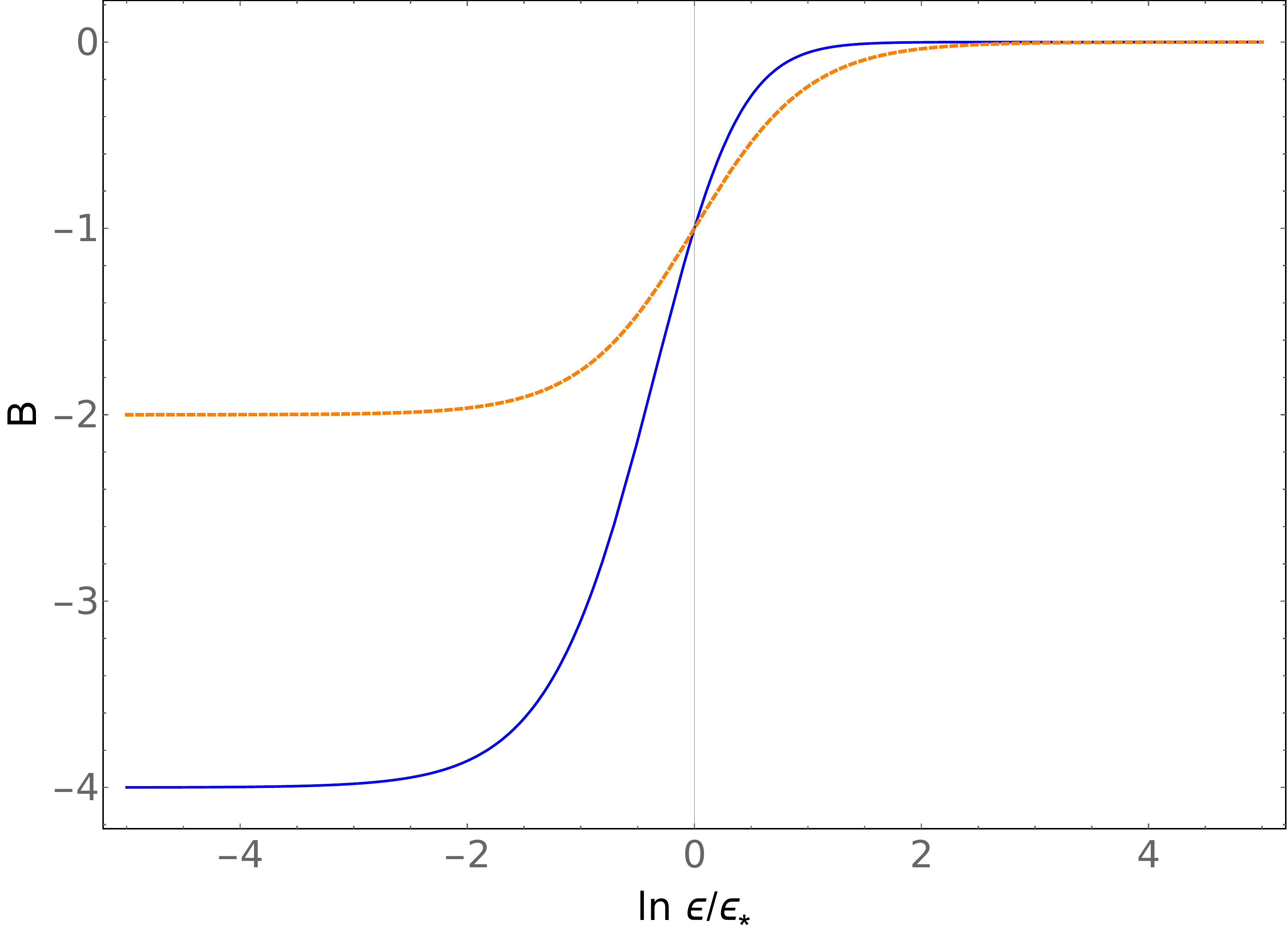} 
\caption{Plot of the RG flow of $B = (Z\alpha)^4(\hat d_s + \hat d_v)$ [solid blue] and $B = (Z\alpha)(\hat c^\prime_s + \hat c^\prime_{v\,\text{tot}})$ [dashed orange] vs $\ln \epsilon/\epsilon_\star$, with $\eta_+ = -1$.} \label{fig:class2} 
\end{center}
\end{figure}
\begin{figure}[ht]
\begin{center}
\includegraphics[width=0.47\textwidth]{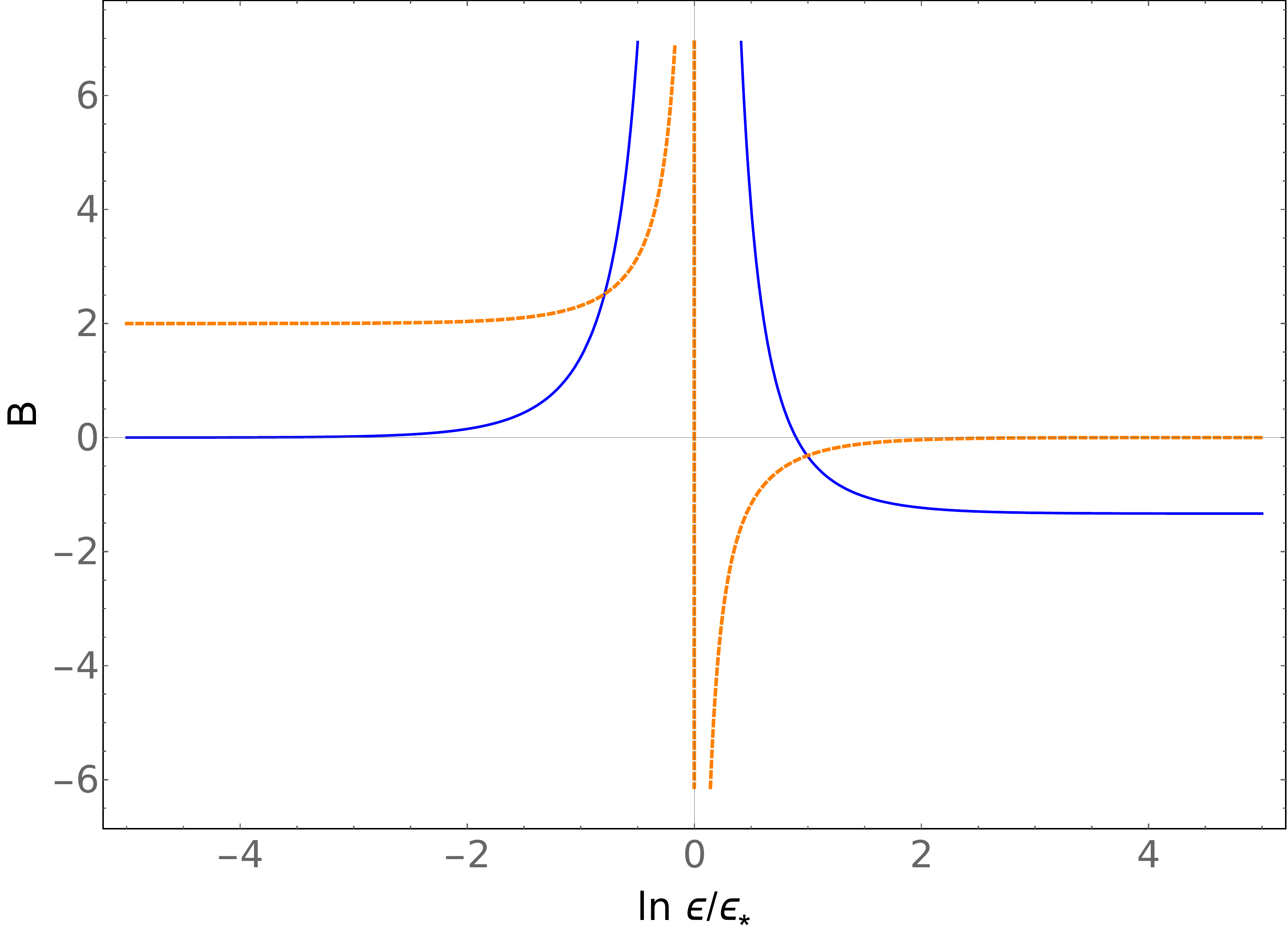} 
\caption{Plot of the RG flow of $B = (Z\alpha)^4(\hat d_s - \hat d_v)$ [solid blue] and $B = (Z\alpha)(\hat c^\prime_s - \hat c^\prime_{v\,\text{tot}})$ [dashed orange] vs $\ln \epsilon/\epsilon^-_\star$, with $\eta_- = +1$.}\label{fig:class3} 
\end{center}
\end{figure}

\begin{figure}[ht]
\begin{center}
\includegraphics[width=0.47\textwidth]{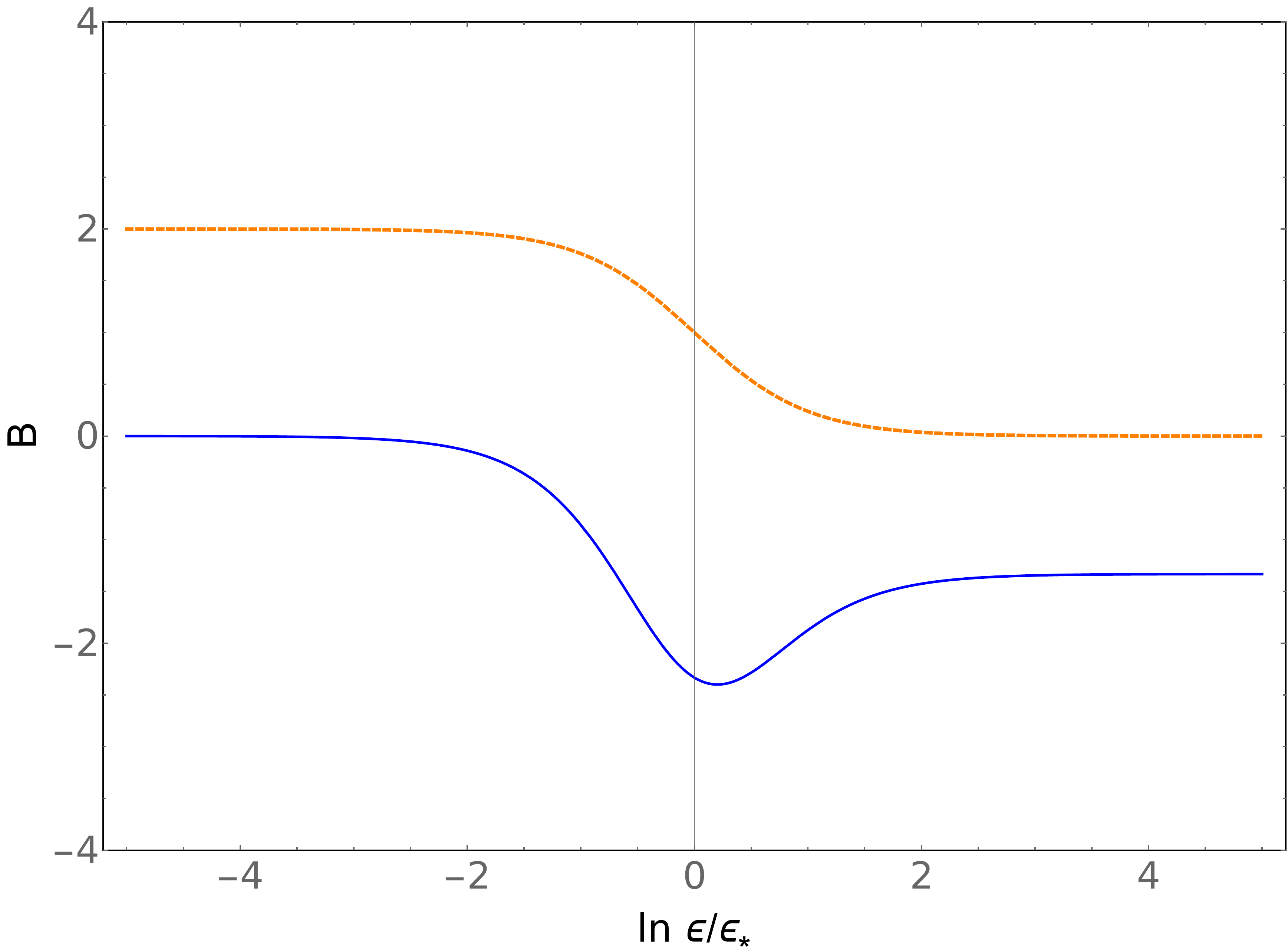} 
\caption{Plot of the RG flow of $B = (Z\alpha)^4(\hat d_s - \hat d_v)/4$ [solid blue] and $B = (Z\alpha)(\hat c^\prime_s - \hat c^\prime_{v\,\text{tot}})$ [dashed orange] vs $\ln \epsilon/\epsilon^-_\star$, with $\eta_- = -1$.} \label{fig:class4} 
\end{center}
\end{figure}

These functions are plotted in Figures \ref{fig:class1} and \ref{fig:class2} (for parity even) and in Figures \ref{fig:class3} and \ref{fig:class4} (for parity odd). In each case, the two classes of flows identified by $\eta_\pm := \text{sgn}(\abs{(Z\alpha)\bar\lambda_\pm -K} - 1)$. These figures show that the RG-invariant quantities $\epsilon^\pm_\star$ give the value of $\epsilon$ for which $(Z\alpha)\bar\lambda_\pm \mp K$ approaches infinity (when $\eta_\pm = +1$) or 0 (when $\eta_\pm = -1$).

\section{Nuclear Uncertainties}
\label{sec:Comps}

Having established in the previous section why the precise value of $\epsilon$ carries no physical information, we turn in this section to connecting the RG-invariant parameters $\epsilon^\pm_\star$ to explicit nuclear properties. 

This is done in the first subsection by computing the energy shift as a function of $\epsilon_\star^\pm$, and then comparing this result to the results of explicit simple models of the nucleus. The upshot of this section is the observation that a single parameter, $\epsilon_\star := \epsilon_\star^+ = \epsilon_\star^-$, accounts for the energy shifts found using explicit calculations with these models, with $\eta_+=\eta_- = +1$. 

Furthermore, the parameter $\epsilon_\star$ required to obtain this agreement does not depend on the quantum numbers $\{n,l,m\}$ of the state whose energy is being computed, as is intuitively plausible given that $\epsilon_\star$ captures the properties of the nucleus and these should not depend on which particular electron (or muon) state that is used to probe them. 

Finally, the above statements are equally true at lowest order and when higher-order contributions are included in powers of $Z\alpha$ and/or $mRZ\alpha$. Working to subdominant order does not introduce new parameters beyond $\epsilon_\star$ into the result, it just determines the value of $\epsilon_\star$ with more precision than at lower order.

The upshot to the order we work is that all calculations are captured by an RG-invariant scale $\epsilon_\star$ of the following form:
\begin{equation}
	\label{eq:epsExp}
	\epsilon^2_\star = (Z\alpha)^2 \left(R_0^2 + R_1^2(Z\alpha) + R_2^2(Z\alpha)^2 + \cdots \right), 
\end{equation}
The length-scales $R_i$ are generalized nuclear moments whose values can weakly depend on $m$ (e.g., logarithmically), and are computed below for several models of interest. Notice in particular that the overall factor $(Z\alpha)^2$ ensures $\epsilon_\star$ is much smaller than the $R_i$, which turn out to be typical nuclear scales.

Finally, the second subsection in this part of the paper asks for observable combinations of energy levels from which $\epsilon_\star$ drops out. Such combinations must always exist when there are more observables than there are nuclear parameters. What is crucial is that the numbers $R_0$, $R_1$, and $R_2$ above are {\it not} independent parameters in this sense, since they enter into all observables --- for {\it both} electronic and muonic atoms --- purely through the single combination $\epsilon_\star$. Because of the explicit appearance of $m$, and the implicit dependence of the $R_i$ on lepton mass, $\epsilon_\star$ will be numerically different between electronic and muonic atoms.  

\subsection{Moments and Polarizabilities}

We start by making contact with nuclear models, computing the value of $\epsilon_\star^\pm$ required to reproduce energy-shift calculations in the literature (and justifying eq.~\eqref{eq:epsExp}).

\subsubsection{RG-invariant Energy Shifts}

Consider first the energy shifts for atomic energy levels as a function of the RG-invariant parameters $\epsilon_\star^\pm$ and $\eta^\pm$. The calculation is greatly simplified given the knowledge that $\epsilon_\star^\pm$ proves to be much smaller than typical nuclear sizes (in retrospect due to the explicit factor $\epsilon_\star \propto Z\alpha$ implied by \eqref{eq:epsExp}). 

Expanding eqs.~\eqref{eq:origRunComb}, \eqref{eq:defBarLam}, \eqref{eq:runD+} and \eqref{eq:runD-} in the limit of small $\epsilon_\star/\epsilon$ --- and specializing to $j=1/2$ --- we have:
 \begin{align}
 	\label{eq:gs}
	\hat g_1 &= -\frac 12 + \frac{2\eta_+}{(Z\alpha)^2}\left( \frac{\epsilon^+_\star}{\epsilon} \right)^2, \notag \\
	\hat g_2 &= -\frac{n^2 - 1}{6n^2} + \frac{4\eta_+}{(Z\alpha)^2}\left( \frac{\epsilon^+_\star}{\epsilon} \right)^2 + \frac{8}{(Z\alpha)^4}\left( \frac{\epsilon^+_\star}{\epsilon} \right)^4, \\
		\hat g_3 &= - \frac 18 -\frac{\eta_+}{(Z\alpha)^2}\left( \frac{\epsilon^+_\star}{\epsilon} \right)^2\left[1 + 2\ln(\epsilon^+_\star/\epsilon)\right] + \frac{2}{(Z\alpha)^4}\left( \frac{\epsilon^+_\star}{\epsilon} \right)^4, \notag
 \end{align}
while
 \begin{equation}
 	\label{eq:fs}
	\begin{split}
	\hat f_1 &= \frac 23, \\
	\hat f_3 &= \frac 12 - \frac{2\eta_-}{(Z\alpha)^2}\left( \frac{\epsilon^-_\star}{\epsilon} \right)^2.
	\end{split}
 \end{equation}
Using these in the energy shifts, eqs.~\eqref{eq:even} and \eqref{eq:odd}, then gives the parity-even $j=1/2$ shift,
\begin{widetext}
\ba \label{eq:betterEvenNew}
	\delta E_{nS_{1/2}} &\simeq& \frac{4m^3 (Z\alpha)^2}{n^3} \eta_+(\epsilon^+_\star)^2 \left\{ 2+ \left[ \frac{12n^2-n-9}{2n^2(n+1)} - 2\ln \left( \frac{2m\epsilon^+_\star Z\alpha}{n} \right) - 2H_{n+1} - 2\gamma + 4\right] (Z\alpha)^2 +\dots\right\}\,,\nn\\
 \ea
\end{widetext}
 while the parity-odd $j = 1/2$ state shifts by
 \begin{equation}
	 \label{eq:betterOddNew}
	 \begin{split}
	 \delta E_{nP_{1/2}}  &\simeq  2\frac{n^2 - 1}{n^5}m^3 (Z\alpha)^4 \eta_-(\epsilon_\star^-)^2\Bigl( 1 + \cdots \Bigr)\,.
	 \end{split}
 \end{equation}
As mentioned earlier, the nuclear shifts to $j = 3/2$ states and higher are smaller than the order to which we work. 

\subsubsection{Fixed Charge Distributions}

The simplest nuclear model treats it as a simple static charge distribution, $\rho(\bfr)$, and energy shifts for Dirac fermions orbiting such distributions have been computed in the limit where the radius, $R$, of the distribution is much smaller than atomic size $a_\ssB$ \cite{Dirac, Structure, Eides, Zemach, Friar, Nickel, NASA}. 

For such models in the limit $R \ll a_\ssB$ the finite-size energy shift to leading and subleading order in $R/a_\ssB$ is parameterized by just three moments of the charge distribution. Expressions for this shift (as found by refs.~\cite{Dirac, Structure, Eides, Zemach, Friar, Nickel, NASA}) agree with \eqref{eq:betterEvenNew} and \eqref{eq:betterOddNew} when $\eta := \eta_+ = \eta_- = +1$ and the RG-invariant parameter $\epsilon_\star^+ = \epsilon_\star^- =: \epsilon_\star$ is given by
\be \label{epsstardist}
\epsilon_\star^2 = \frac{(Z\alpha)^2}{12} \left( r_p^2  + \frac{1}{2}\; r_\ssF^3 mZ\alpha + a^2_{\text{rel}}(Z\alpha)^2 \right) \,,
\ee
which corresponds to the generalized moments
\be \label{eq:epsExpInd}
		R_0^2 = \frac{r_p^2}{12}, \qquad
		R_1^2 = m\frac{r_\ssF^3}{24}, \qquad \hbox{and} \qquad 
		R_2^2 = \frac{a^2_\text{rel}}{12}\,.
\ee
The nuclear moments $r_p^2$, $r_\ssF^3$, and $a^2_\text{rel}$ above are defined as follows. 

At order $m^3R^2(Z\alpha)^4$, the only moment that appears is the charge radius
\be 
\label{rpDef}
 r_p^2 := \frac{1}{Ze} \int \exd^3\bfr \;r^2 \rho \,.
\ee
At order $m^4R^3(Z\alpha)^5$ only the Friar (or third Zemach) moment appears
\be 
\label{eq:friar}
 r_\ssF^3 := \frac{1}{(Ze)^2} \int \exd^3\bfr \exd^3 \bfr'\; \rho(\bfr) \rho(\bfr') | \bfr - \bfr'|^3 \,.
\ee
Finally, at order $m^3R^2(Z\alpha)^6$, there is one more moment that arises which we call $a_\text{rel}$. This moment has a more complicated structure, for which several authors have presented different but equivalent formulations \cite{Friar, Nickel, 3Photon}. Following \cite{3Photon}, we write 
\be
  a^2_\text{rel} = r_p^2 \left[1 + \frac 12 \, \ln(12) - \ln(Z\alpha) +  \ln\left( \frac{r_{\ssC 1}}{r_p} \right) \right]\,,
\ee
with the parameter $r_{\ssC 1}$ ({\it c.f.}~equation (66) in \cite{3Photon}) given by
\begin{equation}
	\begin{split}
		&\ln\frac{r_{\ssC 1}}{r_p} - 1 = \frac{6}{r_p^2}\int_0^\infty \exd r \ln(r/r_p)\dv{r}r^3\\
									  &\quad\times	\qty{2\pi \rho(r)[V^{(2)}(r)]^2 - [V(r)]^2V^{(2)}(r) - \frac{1}{r^2}\qty[\frac r2 + \frac{r_p^2}{6r}]},
	\end{split}
\end{equation}
where $V(r) \approx 1/r$, $V^{(2)}(r) \approx -r/2 - r_p^2/6r$, and $\rho(r)$ is the nuclear charge distribution. 

This example illustrates several things. First it shows that agreement with calculated energy shifts requires the parity-even and parity-odd RG invariants to be the same. This seems a reasonable consequence of the assumed parity-invariance of the nuclear couplings: odd- and even-parity electrons (or muons) see the same nucleus. Furthermore, this example shows how moving past leading order does not introduce new independent RG-invariant parameters into the energy shifts. Instead it provides a more accurate determination of the value of the single RG-invariant $\epsilon_\star$. Finally $\epsilon_\star$ is independent of the lepton-state quantum numbers, $j j_z \Pi$.

\subsubsection{Nuclear Polarizability}

In general, nuclear contributions to atomic energy shifts arise that cannot be simply parameterized in terms of a static nuclear charge distribution, such as those due to ``inelastic'' Coulomb exchanges. These typically involve sums over intermediate nuclear states and so sample nuclear degrees of freedom outside of their ground state, and contain the effects of nuclear polarizability. A representative example of how such a calculation proceeds is sketched in Appendix \ref{app:polariz}. 

The upshot of these calculations is that they contribute (to within the accuracy we work here) to atomic energy shifts in a way that depends on the quantum numbers of the atomic state in the same way as does the charge-radius contribution. As a result these contributions can also be captured by a shift in the value of the RG-invariant scale $\epsilon_\star$ with $\eta = +1$. 

In terms of the parameterization of eq.~\eqref{eq:epsExp} the calculations of refs.~\cite{Friar2,Ji,3Photon} give contributions that first arise at order $m^3R^2(Z\alpha)^5$ for muonic atoms, and $m^4R^3(Z\alpha)^5$ for electronic atoms. For muons, the inelastic two-photon exchange introduces a new contribution $R_1^2 \supset -\tilde \alpha_{\text{pol}}^\prime/6$, where $\tilde \alpha_{\text{pol}}^\prime$ is a generalized (mass-dependent) nuclear polarizability given by \cite{PachuckiPolRev}
	\begin{equation}
		\label{eq:defMuPol}
		\tilde\alpha_{\text{pol}}^\prime := \int_{E_T}\exd E \sqrt{\frac{m}{2E}}\vert\mel{\phi_N}{\vec{d}}{E}\vert^2,
	\end{equation}
	where $\ket{\phi_N}$ is the nuclear ground state, $\ket{E}$ is the nuclear excited state with energy $E-M$, $\vec{d}$ is the nuclear dipole operator (divided by the elementary charge), and $E_T$ is the nuclear threshold excitation energy (which for Helium\cite{BigNuc} is $\sim 20$MeV). 
	Furthermore, at order $m^4R^3(Z\alpha)^5$ the nuclear polarizability also adjusts the value of $R_1^2$, so that $R_1^2 \supset m\tilde r_F^3/24$, where now $\tilde r_F^3$ is a generalized Friar moment. For muonic atoms \cite{Ji}, 
	\begin{equation}
		\tilde r_{F,\mu}^3 = -\frac{1}{(Ze)^2}\int\exd^3 r\int\exd^3 r^\prime |r-r^\prime|^3 \mel{\phi_N}{\hat\rho^\dagger(r)\hat\rho(r^\prime)}{\phi_N},
	\end{equation}
	where $\hat\rho(r)$ is the (un-normalized) nuclear charge density \emph{operator}, and $\ket{\phi_N}$ is again the nuclear ground state (note that the matrix element $\mel{\phi_N}{\hat\rho^\dagger(r)\hat\rho(r^\prime)}{\phi_N}$ is distinct from $\rho(r)\rho(r^\prime) = \mel{\phi_N}{\hat\rho^\dagger(r)}{\phi_N}\mel{\phi_N}{\hat\rho(r^\prime)}{\phi_N}$, which appears in \eqref{eq:friar}). For electronic atoms, the static dipole polarizability also arises at this order, and so \cite{PachuckiPolRev}%
	\begin{align}
		\tilde r_{F,e}^3 &= -\frac{\tilde\alpha_{\text{pol}}}{6} - \frac{1}{(Ze)^2} \\
						 &\quad{}\times\int\exd^3 r\int\exd^3 r^\prime |r-r^\prime|^3 \mel{\phi_N}{\hat\rho^\dagger(r)\hat\rho(r^\prime)}{\phi_N}, \notag
	\end{align}
	where 
	\begin{align}
		\label{eq:def:tildAlphaPol}
		&\tilde\alpha_{\text{pol}} = \frac{2}{3}\int \exd E \qty{\frac{19}{6}\vert\mel{\phi_N}{\vec{d}}{E}\vert^2 +5\vert\mel{\phi_N}{\vec{d}\ln(2E/m)}{E}\vert^2 } \notag \\
														   & \hspace{16em} \text{(muons)} 
	\end{align} 
	is the weighted static electric nuclear polarizability. Finally, $R_2^2$ is also altered, although in this case the exact form of the inelastic exchange is not known for Helium \cite{3Photon}. However, for both electrons and muons it is expected to be well-described by a local interaction due to the high excitation energy of the ${}^4$He nucleus relative to atomic scales, and so should appear as some generalized $a_\text{rel}$ which we denote $\tilde a_\text{rel}$, in analogy with the generalized Friar moment. 
Altogether, inclusion of nuclear polarizability effects can be encoded simply by the contributions 
\begin{equation}
	\label{eq:epsExpIndFull}
		R_0^2 = \frac{r_p^2}{12}\,, \quad
		R_1^2 = -\frac{\tilde \alpha^\prime_\text{pol}}{6} + m\frac{\tilde r_F^3}{24} \quad\hbox{and} \quad
		R_2^2 = \frac{\tilde a_\text{rel}}{12},
\end{equation}
where $\tilde\alpha_\text{pol}^\prime$ is defined in \eqref{eq:def:tildAlphaPol} for muons, and is 0 for electrons.

The bottom line is again that these contributions represent particular kinds of contributions to $\epsilon_\star$, and are not contributing to atomic energy shifts as independent parameters. Consequently assessments of nuclear errors involved in each of these kinds of processes can be interpreted as contributions to the total theoretical uncertainty in microscopic predictions for $\epsilon_\star$.

However, the real power of the above expressions in terms of $\epsilon_\star$ is in their very broad generality. Although specific kinds of nuclear physics contribute to the value of $\epsilon_\star$, the same would also be true for {\it arbitrary} short-distance physics, regardless of this has nuclear origins or not. Because the PPEFT framework parameterizes all possible interactions localized at the nucleus consistent with symmetries, the contribution to atomic energies of these couplings (through their RG-invariant parameterizations $\epsilon_\star$ and $\eta$) are guaranteed to capture any short-distance physics that shares these symmetries to the given order in $R/a_\ssB$ --- regardless of the details of how that physics might be modelled. 

\subsection{Nucleus-Independent Combinations}

Exploitation of more precise measurements of atomic level spacings is currently hampered by theoretical uncertainties associated with predicting the energy shifts due to nuclear physics. Ongoing efforts are underway to improve the theoretical prediction for these nuclear shifts, and in the language of PPEFT these can be regarded as improving the theoretical prediction for the RG-invariant parameter $\epsilon_\star$. In this view the various individual contributions to nuclear level shifts --- {\it e.g.}~charge-radius, Friar moment, polarizability and so on --- all enter together only through this single parameter.\footnote{Since $\epsilon_\star$ depends explicitly on the lepton mass -- {\it c.f.}~the $R_2^3$ term in \eqref{eq:epsExp} -- strictly speaking there is a single parameter controlling electron-type atoms and another one for muonic atoms, and any evidence for a difference in these parameters for electrons and muons is evidence for the presence of a nonzero size for the parameter $R_2^3$.}

The fact that the nucleus can only influence atomic levels through $\epsilon_\star$ suggests another approach towards reducing theoretical error for precision atomic measurements. Rather than trying to reduce the theoretical error by computing this parameter more accurately, why not instead identify combinations of observables from which the parameter $\epsilon_\star$ cancels out? Any such a combination is a quantity for which the theoretical error is much smaller, since it does not depend at all on any nuclear uncertainties. 

To formalize this we write the energy levels of hydrogenic atoms as:
\begin{equation} \label{eq:totShift}
	E_{nj\pm} = E^{\rm Dirac}_{nj} + \delta E^{\text{QED}}_{n j \pm} + \delta E^{\text{PP}}_{n j \pm} + \delta E^{\text{PP QED}}_{n j \pm} \,,
\end{equation}
where quantum numbers $n$, $j$ and parity $\pm$ are used as labels. Here $E^{\rm Dirac}$ is the energy eigenvalue predicted by the Dirac-Coulomb solution, and $\delta E^{\rm QED}$ contains all QED radiative corrections in the limit of a point nucleus. $\delta E^{\rm PP}$ is the nucleus-dependent contribution given above, and $\delta E^{\text{PP QED}}$ contains the influence of nonzero nuclear size on all QED radiative corrections. 

When comparing to the literature --- such as the three-photon contributions evaluated in ref.~\cite{3Photon} --- it is the `high-energy' parts of graphs whose effects can be captured by a shift in the parameters of the effective theory, which in the present instance means shifting the value of $\epsilon_\star^2$ in $\delta E^{\rm PP}_{nj\pm}$. The same cannot be done for the `low-energy' parts that correspond to graphs evaluated within the effective theory using nucleus-modified propagators and so these contributions are either already included in the perturbative expansion of the energy shifts (equations \eqref{eq:even} and \eqref{eq:odd}), or else grouped into $\delta E^{\rm {PP QED}}_{nj\pm}$.

Both of $\delta E^{\rm PP}$ and $\delta E^{\text{PP QED}}$ suffer from systematic uncertainties arising from nuclear physics (and the proton radius problem, should this prove not to be due to experimental error). But because $\delta E^{\text{PP QED}}$ starts out with higher powers of $\alpha$ it only depends on the lowest-order $R_0^2$ contributions\footnote{Apart from logarithms -- see {\it e.g.} \cite{3Photon} -- inasmuch as other nuclear scales besides $R_0$ can appear logarithmically in low-energy contributions. When this occurs we write a contribution of the form $\ln(m R_x)$ as $\ln (m R_0) + \ln(R_x/R_0)$ and absorb the $m$-independent factor $\ln(R_x/R_0)$ into the $R_2^2$ term of \eqref{eq:epsExp}.} to $\epsilon_\star$, unlike $\delta E^{\rm PP}$ which in principle depends on all of the parameters $R_0^2$ through to $R_2^2$ of eq.~\eqref{eq:epsExp}.

However it is differences $\Delta E_{1\to2} := E_{n_1, j_1, \pm_1} - E_{n_2, j_2, \pm_2}$ between energy levels that are measured spectroscopically. For these quantities we therefore write:
\begin{equation} \label{eq:Delta1}
	\Delta E_{1\to2} = \Delta E^{\text{EM}}_{1 \to 2} + \Delta E^{\text{PP}}_{1 \to 2} + \Delta E^{\text{PP QED}}_{1 \to 2} ,
\end{equation}
in which the Dirac-Coulomb and point-source QED effects are grouped together into the term labelled ``EM''.  
Because $\Delta E_{1\to2}^{\rm EM}$ is calculable with negligible error, we focus below on the nucleus-dependent combination
\be
  \widehat{\Delta E}_{1\to2} := \Delta E_{1\to2} - \Delta E^{\text{EM}}_{1 \to 2} = \Delta E^{\text{PP}}_{1 \to 2} + \Delta E^{\text{PP QED}}_{1 \to 2} \,.
\ee

Our goal is to identify linear combinations of these observables from which the parameter $\epsilon_\star$ cancels. With upcoming experiments in mind we do so explicitly here for muonic atoms up to the accuracy of $m^4R^3 (Z\alpha)^5$ required to see the Friar moment. For electrons we go to the same accuracy, which is slightly more involved due to the necessity of keeping terms at both order $m^4R^3 (Z\alpha)^5$ and $m^3R^2 (Z\alpha)^6$, since these are similar in size (due to the numerical coincidence $m_e R \sim Z\alpha$). 

\subsubsection{Predicted Energy Differences}

In order to pursue this program we need complete expressions for the $\epsilon_\star$ dependence of all relevant levels, including both the $\delta E^{\rm PP}$ and $\delta E^{\text{PP QED}}$ contributions. Since to the desired accuracy $\epsilon_\star$ does not appear at all within $\delta E^{\rm PP}$ for the energies of $j > \frac12$ states, we focus on itemizing all relevant contributions for $j =\frac12$. 

The mixed nuclear-QED contribution has been evaluated at the order required, and we simply quote the result here. For both electrons and muons the leading result is given by \cite{QEDPPe, QEDPPmu} 
\ba
  \delta E^{\text{PP QED}} &=& \frac{4\eta^{(e)}_{nl}}{n^3} \,m_\mu^3 \alpha (Z\alpha)^2 \epsilon_{\star \mu}^2 \qquad \hbox{(muons)} \nn\\
  &=& \frac{4 \eta^{(\mu)}_{nl}}{n^3} \,m_e^3 \alpha (Z\alpha)^3 \epsilon_{\star e}^2 \qquad \hbox{(electrons)}\,, 
\ea
with the dimensionless coefficients $\eta^{(a)}_{nl}$ depending on the quantum numbers of the lepton state. In these expressions the subscripts `$a = \mu\,, e$' on $\epsilon_{\star \mu}$ is meant to underline that it is evaluated using the muon mass in its $R_2^3 m Z\alpha$ contribution.  We identify the $\epsilon_\star$-dependence by trading the dependence on $r_p^2$ given in the literature for $\epsilon_\star$ using only the leading, $R_0^2$, contribution from eq.~\eqref{epsstardist}: $\epsilon_\star^2 = \frac{1}{12} \, (Z\alpha)^2  r_p^2$.

For electronic atoms $\eta_{nl}$ is given by \cite{QEDPPe}
\begin{equation}
	\label{eq:EtaE}
	\eta^{(e)}_{nl} := \left( 8\ln 2 - 10 \right) \delta_{l0} \quad \text{(electron)} \,,
\end{equation}
which vanishes for $l \neq 0$ since the wave-function must have support at the position of the nucleus because the Bohr radius for the orbit, $a_\ssB \sim (Z \alpha m_e)^{-1}$, is much larger than the Compton wavelength, $\lambda_c \sim m_e^{-1}$, of the virtual electrons in the QED loop. The same is not true for muons since $\alpha m_\mu$ is comparable to $m_e$, and so for muonic atoms $\eta^{(\mu)}_{nl}$ need not vanish for $l \neq 0$. The precise value of $\eta^{(\mu)}_{nl}$ --- given in \cite{QEDPPmu} --- is not required in what follows.

Collecting results we have:

\medskip\noindent
\textbf{Muons}: Here we have the non-zero nuclear-dependent energy shifts to the desired order\footnote{We switch to spectroscopic notation where states are labelled by $j$ and parity, so the labels $S,P,D,F,\cdots$ are proxies for parity. Thus $S$ (or $P$) are parity-even (-odd) states with spin $j = \frac12$, while $D$ (or $F$) are parity-even (-odd) with spin $j = \frac32$ and so on.}
		\begin{equation}
			\delta E_{nS_{1/2}}^{\rm PP} + \delta E_{nS_{1/2}}^{\rm PP QED}  = \frac{4m_\mu^3 (Z\alpha)^2}{n^3}(2 + \eta^{(\mu)}_{n0}\, \alpha )\epsilon^2_{\star\mu}
		\end{equation}
		while 
		\begin{equation}
			\delta E_{nP_{1/2}}^{\rm PP} + \delta E_{nP_{1/2}}^{\rm PP QED} = m_\mu^3 (Z\alpha)^2 \eta^{(\mu)}_{n1} \, \alpha \, \epsilon^2_{\star\mu},
		\end{equation}
		and 
		\begin{equation}
			\delta E_{nP_{3/2}}^{\rm PP}+\delta E_{nP_{3/2}}^{\rm PP QED} = m_\mu^3 (Z\alpha)^2 \eta^{(\mu)}_{n1} \, \alpha \, \epsilon^2_{\star\mu}.
		\end{equation}

		Combining these expressions provides the following expressions for the measurable energy differences for the lowest angular momentum states:
\begin{equation}
			\label{eq:epsShift1}
			\widehat{\Delta E}_{nS_{1/2} - nP_{1/2}}   = \frac{4m_\mu^3 (Z\alpha)^2}{n^3}\Bigl[ 2 + \left(\eta^{(\mu)}_{n0} - \eta^{(\mu)}_{n1} \right)\,  \alpha \Bigr] \epsilon^2_{\star\mu} \,,
		\end{equation}
				\begin{equation}
			\label{eq:epsShift1a}
			 \widehat{\Delta E}_{nS_{1/2} - nP_{3/2}}   = \frac{4m_\mu^3 (Z\alpha)^2}{n^3}\Bigl[ 2 +\left(\eta^{(\mu)}_{n0} - \eta^{(\mu)}_{n1}\right)\,  \alpha \Bigr] \epsilon^2_{\star\mu} \,,
		\end{equation}
		while
		\begin{equation}
			\label{eq:epsShift2}
			\widehat{\Delta E}_{nP_{1/2} - nP_{3/2}}  = 0 \,.
		\end{equation}

\medskip\noindent
 \textbf{Electrons}:  The corresponding formulae for electrons are
 \begin{equation}
 	\begin{split}
			&\delta E_{nS_{1/2}}^{\rm PP} + \delta E_{nS_{1/2}}^{\rm PP QED}  = \frac{4m_e^3 (Z\alpha)^2}{n^3} \epsilon^2_{\star e}\left\{2 + \left[\frac{12n^2-n-9}{2n^2(n+1)}  \right. \right. \\
			& \left.\left. {} - 2\ln \left( \frac{2m_e\epsilon_{\star e}Z\alpha}{n} \right) - 2H_{n+1} - 2\gamma + 4 + \frac{\eta^{(e)}_{n0}}{Z}\right] (Z\alpha)^2  \right\}\,,
 	\end{split}
 \end{equation}
		as well as
		\begin{equation}
			\delta E_{nP_{1/2}}^{\rm PP} + \delta E_{nP_{1/2}}^{\rm PP QED} = 2\left( \frac{n^2 - 1}{n^5}\right) m_e^3(Z\alpha)^4\epsilon_{\star e}^2,
		\end{equation}
		but now $\delta E_{nP_{3/2}}^{\rm PP} = \delta E_{nP_{3/2}}^{\rm PP QED}  = 0$ to the order of interest.
		
		The corresponding energy differences for electrons are therefore
		\begin{equation}
			\begin{split}
		\label{eq:epsShift3}
		&\widehat{\Delta E}_{nS_{1/2} - nP_{3/2}} =  
		 \frac{4m_e^3 (Z\alpha)^2}{n^3} \epsilon^2_{\star e}\left\{2 + \left[\frac{12n^2-n-9}{2n^2(n+1)} \right.\right. \\
		 & \left. \left. - 2\ln \left( \frac{2m_e\epsilon_{\star e}Z\alpha}{n} \right) - 2H_{n+1} - 2\gamma + 4 + \frac{\eta^{(e)}_{n0}}{Z}\right] (Z\alpha)^2 \right\}\,, 
			\end{split}
		\end{equation}
		as well as
			\begin{align}
		\label{eq:epsShift4}
			\widehat{\Delta E}_{nS_{1/2} - nP_{1/2}} &= \frac{4m_e^3 (Z\alpha)^2}{n^3} \epsilon^2_{\star e} \left\{2 + \left[\frac{12n^2-n-9}{2n^2(n+1)} \right. \right. \notag \\
			& \, \quad - 2\ln \left( \frac{2m_e\epsilon_{\star e}Z\alpha}{n} \right) - 2H_{n+1} \\
			& \,\,\,\quad \left.\left. - 2\gamma + 4 + \frac{\eta^{(e)}_{n0}}{Z} - \frac{n^2 - 1}{2n^2}\right] (Z\alpha)^2 \right\}\,, \notag 
			\end{align}
		and 
		\begin{equation}
			\label{eq:epsShift5}
			\widehat{\Delta E}_{nP_{1/2} - nP_{3/2}}  = 2\left( \frac{n^2 - 1}{n^5} \right) m_e^3(Z\alpha)^4\epsilon_{\star e}^2 \,.
		\end{equation}
		
In essence, these expressions imply that the nuclear-size contributions to a great many energy electronic and muonic levels can be parameterized in terms of just two parameters, $\epsilon_{\star e}$ and $\epsilon_{\star\mu}$. By eliminating these parameters we can derive relations that directly connect measurable quantities. The relations derived in this way are therefore known with smaller theoretical errors, since they are entirely independent of nuclear uncertainties.

\subsubsection{Levels with \texorpdfstring{$n = 2$}{n=2}}

We start by concentrating on the energy levels that have already been measured, and so restrict our attention to the special case $n=2$. 

Focussing first on muonic atoms, the nuclear contribution to the differences between the three levels $2S_{1/2}$, $2P_{1/2}$ and $2P_{3/2}$ is controlled by the single parameter $\epsilon_{\star\mu}$. This means there must be a nucleus-independent combination relating the two independent energy differences. This can be taken to be \eqref{eq:epsShift2}: $\widehat{\Delta E}_{nP_{1/2}-nP_{3/2}} = 0$ is a statement unclouded by nuclear uncertainties, in particular for $n =2$.

Alternatively, \eqref{eq:epsShift1} provides an accurate experimental determination of $\epsilon_\star$ for muonic Helium:
\begin{equation}\label{eq:solvedEpsShift1a}
			\epsilon^2_{\star\mu} = \frac{ \widehat{\Delta E}_{2S_{1/2} - 2P_{1/2}} }{ m_\mu^3 (Z\alpha)^2 \left[ 1 + \frac12 (\eta^{(\mu)}_{20} - \eta^{(\mu)}_{21} ) \alpha \right]} + \cO[(Z\alpha)^4]
 \,.
		\end{equation}

Turning now to the ${}^4$He${}^+$ ion, the nuclear contribution to the two independent differences between the $2S_{1/2}$, $2P_{1/2}$ and $2P_{3/2}$ levels is controlled by the single parameter $\epsilon_{\star e}$, again suggesting there is a nucleus-independent combination. 

This can be found by using \eqref{eq:epsShift4} to eliminate $\epsilon_{\star e}$ and using the result in \eqref{eq:epsShift5} to predict the $2P_{1/2}-2P_{3/2}$ transition in terms of the $2S_{1/2}-2P_{1/2}$ transition:
		\begin{equation}
			\label{eq:goodishPredict1}
			\widehat{\Delta E}_{2P_{1/2} - 2P_{3/2}} = \frac{3}{16}(Z\alpha)^2 \widehat{\Delta E}_{2S_{1/2}-2P_{1/2}} + \cO[(Z\alpha)^7].
		\end{equation}
		We write the error in this expression as $(Z\alpha)^7$ rather than $(Z\alpha)^8$ because the corrections to \eqref{eq:epsShift5} arise at relative order $(mRZ\alpha)$, though for electrons this is numerically closer to order $(Z\alpha)^8$. Alternatively, using the $2S_{1/2} - 2P_{1/2}$ to predict the $2S_{1/2} - 2P_{3/2}$ difference leads to the equivalent prediction
		\begin{equation}
			\label{eq:subP3}
			\begin{split}
				\widehat{\Delta E}_{2S_{1/2}-2P_{3/2}} &= \widehat{\Delta E}_{2S_{1/2}-2P_{1/2}}\left[1 + \frac{3}{16} \, (Z\alpha)^2\right. \\
													   &\qquad\qquad\qquad\qquad + \order{(Z\alpha)^4} \Big] \,.
			\end{split}
		\end{equation}

While naively $\epsilon_{\star e}$ might be obtained from \eqref{eq:epsShift5}, leading to
		\begin{equation}
			\label{eq:goodishIsol}
			\epsilon_{\star e}^2 = \frac{16}{3m_e^3(Z\alpha)^4} \widehat{\Delta E}_{2P_{1/2}-2P_{3/2}} + \order{(Z\alpha)^4} \,,
		\end{equation}
		%
this determines it with larger relative error than it would have been by solving for $\epsilon_{\star e}$ from one of the other two energy differences. Taking this latter approach instead leads (see Appendix \ref{app:Astounding}, including the result for general $n$) to
\begin{align}\label{eq:bigSol}
	m_e^2\epsilon_{\star e}^2 &\simeq 
	 \frac{1}{m_e(Z\alpha)^2}\widehat{\Delta E}_{2S_{1/2}-2P_{1/2}}\Big\{ 1  \notag \\
	 &\left\{ +\frac{ (Z\alpha)^2}{2}\left[  \ln(\frac{\widehat{\Delta E}_{2S_{1/2}-2P_{1/2}}}{m_e} ) - \frac{3}{2} + 2\gamma  - \frac{\eta_{20}}{Z} \right]\right\} \notag \\
	 &\qquad\qquad{}+ \order{(Z\alpha)^2\frac{\widehat{\Delta E}_{2S_{1/2}-2P_{1/2}}}{m_e}},
\end{align}
%
%
%
%
%
and the correction is now down by $(Z\alpha)^4$ relative to the leading term.
	
\subsubsection{More general \texorpdfstring{$n$}{n}}

The relations found above for the special case $n=2$ might not be all that surprising. However should experiments access transitions with higher $n$, the fact that all nuclear contributions are controlled by the single parameter $\epsilon_\star$ becomes ever more predictive.  This section makes a start at some of the nuclear-free relations that can be derived in this way for general $n$.

\medskip\noindent
 \textbf{Muons}:  We start with muons, which are simpler. A start is the prediction for the general $nS_{1/2}-nP_{1/2}$ shift for any $n$ given measurements of this shift for $n=2$. For $n=2$, we use \eqref{eq:epsShift1a} to infer the value of $\epsilon_{\star \mu}$, which when substituted into \eqref{eq:epsShift1} for general $n$, gives:
\begin{equation}\label{eq:muPredict1}
    \widehat{\Delta E}_{nS_{1/2}-nP_{1/2}} = \frac{8[2 + \alpha(\eta^{(\mu)}_{n0}-\eta^{(\mu)}_{n1})]}{n^3[2 + \alpha(\eta^{(\mu)}_{20}-\eta^{(\mu)}_{21})]}\; \widehat{\Delta E}_{2S_{1/2}-2P_{1/2}}.
\end{equation}
Similarly, generic muonic $S-S$ transitions become
\begin{equation}\label{eq:muPredict2}
	\widehat{\Delta E}_{n_1S_{1/2}-n_2S_{1/2}} = 2\left(\frac{2 + \alpha\eta^{(\mu)}_{n_1 0}}{n_1^2} - \frac{2 + \alpha\eta^{(\mu)}_{n_2 0}}{n_2^2}\right)\widehat{\Delta E}_{2S_{1/2}-2P_{1/2}}.
\end{equation}
A similar argument relates the $S_{1/2}-P_{3/2}$ transitions for general $n$:
\begin{equation}
	\begin{split}
		&\frac{n_1^2}{2 + \alpha(\eta^{(\mu)}_{n_1 0} - \eta^{(\mu)}_{n_2 1})}\widehat{\Delta E}_{n_1S_{1/2} - n_1P_{3/2}} = \\
		&\qquad\qquad\quad \frac{n_2^2}{2 + \alpha(\eta^{(\mu)}_{n_2 0} - \eta^{(\mu)}_{n_2 1})}\widehat{\Delta E}_{n_2S_{1/2} - n_2P_{3/2}} \,.
	\end{split}
\end{equation}

\medskip\noindent
 \textbf{Electrons}:  
Similar expressions hold for electronic atoms. The prediction for the $nS_{1/2}-nP_{1/2}$ shift for any $n$ in terms of this shift for $n=2$ obtained by using \eqref{eq:bigSol} to infer the value of $\epsilon_{\star e}$ used in \eqref{eq:epsShift4} gives:
\begin{align}
	\label{eq:bigPredict1}
		\widehat{\Delta E}_{nS_{1/2}-nP_{1/2}}	 &= \frac{8}{n^3}\widehat{\Delta E}_{2s_{1/2}-2p_{1/2}} \times\\
												 &\quad\left\{1 + (Z\alpha)^2 \left[N(n) - \frac{n^2 - 1}{4n^2}\right] \right\} \notag\,, 
\end{align}
in which we define
\begin{equation}
	\label{eq:defN}
	N(n) := \frac{12n^2-n-9}{4n^2(n+1)}  - H_{n+1} + \frac 54 + \frac{\eta^{(e)}_{n0}}{2Z} - \frac{\eta^{(e)}_{20}}{2Z} - \ln(\frac{2}{n})\,.
\end{equation}
The predictions for electronic $S-S$ transitions is similarly:
\begin{align}
	\label{eq:bigPredict2}
	\widehat{\Delta E}_{n_1 S - n_2 S} &= \widehat{\Delta E}_{2S_{1/2}-2P_{1/2}}\left\{ \frac{1}{n_1^3} - \frac{1}{n_2^3} \right. \\
									   &\qquad\qquad\left.+ (Z\alpha)^2\left[\frac{N(n_1)}{n_1^3} - \frac{N(n_2)}{n_2^3} \right]\right\}.\notag
\end{align}
The nucleus-free prediction for the difference between the $P$ states for electronic atoms becomes
\begin{equation}\label{eq:linCom1}
    \frac{n_1^5}{n_1^2 - 1} \widehat{\Delta E}_{n_1P_{1/2}-n_1P_{3/2}} = \frac{n_2^5}{n_2^2 - 1} \widehat{\Delta E}_{n_2P_{1/2}-n_2P_{3/2}} \,.
\end{equation}

This situation is somewhat more complicated for electronic $S$-wave states, but using
\begin{align}
		&n_1^3 \widehat{\Delta E}_{n_1S_{1/2}-n_1P_{1/2}}  \\
		&\qquad - n_2^3 \widehat{\Delta E}_{n_2S_{1/2}-n_2P_{1/2}} = 4m_e^3(Z\alpha)^4\epsilon^2_{\star e} (F[n_1]-F[n_2]), \notag
\end{align}
where 
\begin{equation}
	\label{eq:defF}
	F[n]:= \frac{12n^2 - n - 9}{2n^2(n+1)} - \frac{n^2 - 1}{24n^2} + 2\ln n - 2H_{n+1} + \frac{\eta^{(e)}_{n0}}{Z}\,,
\end{equation}
the difference becomes
\begin{align} \label{eq:linCom2}
	&\frac{1}{F[n_1]-F[n_2]}\left(n_1^3\widehat{\Delta E}_{n_1S_{1/2}-n_1P_{1/2}} - n_2^3 \widehat{\Delta E}_{n_2S_{1/2}-n_2P_{1/2}}\right) \notag \\
	&\qquad\qquad\qquad- \frac{24n_1^5}{n_1^2 - 1}\widehat{\Delta E}_{n_1P_{1/2}-n_1P_{3/2}} = 0\,,
\end{align}
which is again free of nuclear uncertainties. It is clear that a great many such relations can be derived in the same way.

\section{Numerical Example}
\label{sec:Nums}

At the moment, data\cite{HeLambExp} is only available for the $2P_{1/2}-2S_{1/2}$ transition in $^4$He$^+$. With this transition, we can use \eqref{eq:bigPredict2} to predict the $1S-2S$ transition in hydrogenic helium, which is relevant for upcoming experiments \cite{HeStoSExp}. Subtracting the point-like physics listed in \cite{Yerokhin}, we compute
\begin{equation}
	\label{eq:expHeHat}
	\widehat{\Delta E}^{\text{(exp)}}_{2S_{1/2} - 2P_{1/2}} = -2.58\, (5) \times 10^{-9}\, \text{Ry}\,,
\end{equation}
in units of the Rydberg energy. Here, the number in parentheses is the error on the last digit. The predicted $1S-2S$ transition is then 
\begin{align}
	\label{eq:predictStoS}
	\Delta E_{2S - 1S} &=  2.9997067118\, (4)\, \text{Ry}\, \notag \\
					   &= 9.868561009\, (1) \times 10^{9} \, \text{MHz},
\end{align}
where in the last line, we used $\text{Ry} = 3.289841960355 \times 10^{15}\, \text{Hz}$ from the 2014 CODATA review \cite{QEDShifts}. Our prediction agrees with \cite{HeStoSExp} and \cite{Yerokhin}, however the error we report is nominally a few times larger than they report (3 times \cite{HeStoSExp} and 4 times \cite{Yerokhin}). What is important in our case is that the error is completely independent of nuclear uncertainties, and is dominated by the experimental error. Our result will therefore only improve as future experiments improve their precision, and will never be hindered by a particular choice of nuclear model.

\section{Conclusion}
\label{sec:conc}

We here apply the PPEFT framework to muonic and electronic atoms with spinless nuclei, which produce systematic parameterizations of the energy level shifts due to \emph{all} short-range physics, including (but not limited to) uncertainties in evaluating nuclear contributions. Our parameterization cleanly identifies a single mass-dependent length-scale, $\epsilon_\star$, that encodes the effect of all nuclear physics on atomic energy levels. 

That is, in discussions of finite-size contributions to atomic energy shifts, one often writes (see e.g.~ \cite{Ji}):
 \begin{equation} \label{eq:yoozh}
	\Delta E = \delta_{\text{QED}} + \delta_{\text{FS}\,r^2}\ev{r_p^2} + \delta_{\text{FS Other}},
 \end{equation}
where $\delta_{\text{QED}}$ is all of the non-finite-size dependent contributions, $\delta_{\text{FS}\,r^2}\ev{r_p^2}$ is all the finite-size terms that are proportional to the squared charge radius, and $\delta_{\text{FS Other}}$ is all the other finite-size contributions. Our observation is that at the level of atomic energy shifts, this division is artificial. The real division is 
\begin{equation} \label{eq:us}
 \Delta E = \delta_{\text{QED}} + \delta_{\epsilon_\star},
\end{equation}
where $\delta_{\text{QED}}$ is all point-nucleus contributions (as above), and $\delta_{\epsilon_\star}$ is all finite-size contributions, which is a known function of the one length scale $\epsilon_\star$. The separation of $\epsilon_\star$ into different sources (such as moments of the nuclear charge distribution, and nuclear polarizability) is a theoretical exercise (although certainly a worthy one) that always needs supplementary information, such as input from theoretical models and scattering data. However, $\epsilon_\star$ is just one number, so once it is determined from a single measurement, it can be used to predict the finite-size contribution of all other measurements.

As a practical application of this observation, we use two different strategies to make predictions about spectroscopic transition for electronic and muonic atoms that are free of $\epsilon_\star$. For these observables our formulae reduce the theoretical error in tests of QED by eliminating any uncertainties arising from explicit models of the nucleus. The same predictions are also independent of any potential short-range new physics (should this prove to be responsible for the proton-radius problem) allowing tests of QED using only muonic ${}^4$He whose validity is undiminished by the existence of such forces.

Our first strategy is using a single measurement to solve for $\epsilon_\star$, and then use that to predict all other measurements. Doing so, we find explicitly predictions for the following transitions:
For muonic atoms, we find
\begin{equation}\label{rec:muPredict1}
    \widehat{\Delta E}_{nS_{1/2}-nP_{1/2}} = \frac{8[2 + \alpha(\eta^{(\mu)}_{n0}-\eta^{(\mu)}_{n1})]}{n^3[2 + \alpha(\eta^{(\mu)}_{20}-\eta^{(\mu)}_{21})]}\; \widehat{\Delta E}_{2S_{1/2}-2P_{1/2}}.
\end{equation}
and
\begin{equation}\label{rec:muPredict2}
	\widehat{\Delta E}_{n_1S_{1/2}-n_2S_{1/2}} = 2\left(\frac{2 + \alpha\eta^{(\mu)}_{n_1 0}}{n_1^2} - \frac{2 + \alpha\eta^{(\mu)}_{n_2 0}}{n_2^2}\right)\widehat{\Delta E}_{2S_{1/2}-2P_{1/2}}.
\end{equation}
while electronic atoms produce
\begin{align}
		\widehat{\Delta E}_{nS_{1/2}-nP_{1/2}}	 &= \frac{8}{n^3}\widehat{\Delta E}_{2s_{1/2}-2p_{1/2}}\times \\
												 &\qquad\left\{1 + (Z\alpha)^2 \left[N(n) - \frac{n^2 - 1}{4n^2}\right] \right\} \,, \notag
\end{align}
and
\begin{align}
	\widehat{\Delta E}_{n_1 S - n_2 S} &= \widehat{\Delta E}_{2S_{1/2}-2P_{1/2}}\left\{ \frac{1}{n_1^3} - \frac{1}{n_2^3} \right. \\
									   &\qquad \left. + (Z\alpha)^2\left[\frac{N(n_1)}{n_1^3} - \frac{N(n_2)}{n_2^3} \right]\right\}, \notag
\end{align}
with $N(n)$ defined in \eqref{eq:defN}.

Our second approach is to avoid solving for $\epsilon_\star$ altogether, and instead find general linear combinations of measurements for which it falls out. In this way, we predict:
For muonic systems,
\begin{equation}
	\begin{split}
		&\frac{n_1^2}{2 + \alpha(\eta^{(\mu)}_{n_1 0} - \eta^{(\mu)}_{n_2 1})}\widehat{\Delta E}_{n_1S_{1/2} - n_1P_{3/2}} = \\
		&\qquad\qquad\quad\frac{n_2^2}{2 + \alpha(\eta^{(\mu)}_{n_2 0} - \eta^{(\mu)}_{n_2 1})}\widehat{\Delta E}_{n_2S_{1/2} - n_2P_{3/2}} \,,
	\end{split}
\end{equation}
while for electronic systems, 
\begin{align} 
	&\frac{1}{F[n_1]-F[n_2]}\left(n_1^3\Delta E_{n_1S_{1/2}-n_1P_{1/2}} - n_2^3 \Delta E_{n_2S_{1/2}-n_2P_{1/2}}\right) \notag \\
	&\qquad\qquad\qquad\qquad- \frac{24n_1^5}{n_1^2 - 1}\Delta E_{n_1P_{1/2}-n_1P_{3/2}} = 0
\end{align}
where $F[n]$ is defined in \eqref{eq:defF}.

Using the only available data for the helium ion (the $2S-2P$ Lamb shift in ordinary $^4$He$^+$), we use \eqref{eq:bigPredict2} to predict a $1S-2S$ transition $\nu_{1S-2S} = 9.868561009\, (1) \times 10^{9} \, \text{MHz}$. While our uncertainty in this prediction is roughly 4 times the uncertainty in the literature, our error is dominated by the experimental precision of the $2S-2P$ measurement. Consequently, our predictions will become more and more precise as experiments improve, and remain unencumbered by the inherent uncertainty in choice of nuclear model.

Although we here address only spinless nuclei, it is certainly possible to include nuclei with spin in the PPEFT framework, and work is ongoing to do so. Though nuclear spin changes the counting of parameters in the energy shift formulae above, the principle remains exactly the same and we expect in this case also to be able to build observables from which short-range contributions completely drop out.

\begin{acknowledgments}
	\label{acknowledgments}
	We thank Richard Hill, Bob Holdom, Marko Horbatsch, Ted Jacobson, Roman Koniuk, Bernie Nickel, Sasha Penin, Maxim Pospelov, Ira Rothstein, Kai Zuber, and Krzysztof Pachucki for discussions and Aldo Antognini, Franz Kottman and Randolf Pohl for very helpful correspondence.  

We thank the organizers of the workshop `Precision Measurements and Fundamental Physics: the Proton Radius and Beyond' held at the Mainz Institute for Theoretical Physics (MITP), for providing such stimulating environs where part of this work was completed.

This research was supported in part by funds from the Natural Sciences and Engineering Research Council (NSERC) of Canada. Research at the Perimeter Institute is supported in part by the Government of Canada through Industry Canada, and by the Province of Ontario through the Ministry of Research and Information (MRI).

\end{acknowledgments}

\appendix

\section{Polarizability in a Nuclear Model}
\label{app:polariz}

To illustrate how nuclear polarizabilities enter into the PPEFT framework this appendix considers a relatively simple nuclear model, following refs.~\cite{Friar2} and \cite{Ji}. The model works with nucleons and leptons with states $\ket{NJJ_z;njj_z}$ representing the nuclear (upper case) quantum numbers and lepton (lower case) quantum numbers. The Hamiltonian of the system is:
\begin{equation}
	H = H_N + H_f + \Delta H
\end{equation}
where $H_N$ is the Hamiltonian of the nucleus (whose details never need be explicitly written, with the basis of nuclear states $\ket{NJJ_z}$ assumed known), $H_f$ is the Schrodinger \emph{or} Dirac Coulomb Hamiltonian for the lepton interacting with a point-source Coulomb potential, and $\Delta H$ is given by:
\begin{equation}
	\Delta H = \frac{Z\alpha}{r} - Z\alpha\int \exd^3 r^\prime \frac{\hat\varrho(\mathbf{r}^\prime)}{\abs{\mathbf{r} - \mathbf{r}^\prime}}
\end{equation}
where $\hat \varrho$ is the electric charge operator written in terms of the quantum nuclear degrees of freedom (such as the nucleon positions and charges). The perturbation subtracts out the point-source Coulomb interaction appearing in $H_f$ and replaces it with the more realistic nuclear electromagnetic source distribution.  

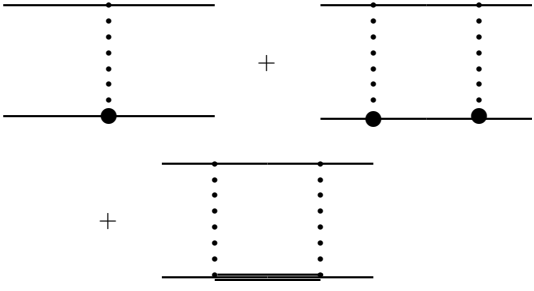
\begin{figure}
\centerline{
\begin{picture}(260,110)
\put(0,40){\begin{picture}(100,100)
    \thicklines
    \multiput(64,20)(0,6){8}{\circle*{2}}
    \put(64,20){\line(1,0){40}}
    \put(64,20){\line(-1,0){40}}
    \put(64,62){\line(1,0){40}}
    \put(64,62){\line(-1,0){40}}
    \put(64,20){\circle*{6}}
    \end{picture}}
\put(110,40){\begin{picture}(160,100)
    \put(10,38){+}
    \end{picture}}
\put(120,40){\begin{picture}(100,100)
    \thicklines
    \multiput(44,20)(0,6){8}{\circle*{2}}
    \multiput(84,20)(0,6){8}{\circle*{2}}
    \put(64,19){\line(1,0){40}}
    \put(64,19){\line(-1,0){40}}
    \put(64,62){\line(1,0){40}}
    \put(64,62){\line(-1,0){40}}
    \put(44,19){\circle*{6}}
    \put(84,20){\circle*{6}}
    \end{picture}}
\put(50,-20){\begin{picture}(160,100)
    \put(10,38){+}
    \end{picture}}
\put(60,-20){\begin{picture}(100,100)
    \thicklines
    \multiput(44,20)(0,6){8}{\circle*{2}}
    \multiput(84,20)(0,6){8}{\circle*{2}}
    \put(64,19){\line(1,0){40}}
    \put(84,18){\line(-1,0){40}}
    \put(85,20){\line(-1,0){40}}
    \put(64,19){\line(-1,0){40}}
    \put(64,62){\line(1,0){40}}
    \put(64,62){\line(-1,0){40}}
    \end{picture}}
\end{picture}
}
\caption{Graphs arising from the perturbative expansion of nuclear electromagnetic interactions. The upper solid line represents the atomic lepton, dotted lines represent the Coulomb part of the electromagnetic field and the lower solid line represents the nucleus in its ground state. The fat dot vertex includes a non-pointlike momentum-dependent correction to the Coulomb vertex, while the fat solid line represents the propagation of an excited nuclear state. Crossed graphs are also present even though they are not drawn. \label{figure:PhotonExchange}}
\end{figure}

Working perturbatively in $\Delta H$ leads to a graphical expansion that includes those of Fig.~\ref{figure:PhotonExchange}. Of these, the left-most graph is linear in the non-pointlike Coulomb-nuclear coupling $\Delta H$, and involves one factor of the nuclear charge-density operator $\hat \varrho$ evaluated within the nuclear ground state $\mel{0}{\Delta H}{0}$. It is this type of graph that gives the contributions that look like the charge-radius $r_p^2$ of the nuclear charge distribution $\rho(\bfr) = \mel{0}{\hat\varrho(\bfr)}{0}$.

Terms quadratic in this same nuclear charge distribution, such as the Friar moment $r_\ssF^3$ of \eqref{eq:friar}, arise from the second graph in Fig.~\ref{figure:PhotonExchange} that are quadratic in $\Delta H$ but also only involve the nucleus in its ground state. The first two types of graphs therefore have counterparts for leptons interacting with a specified charge distribution and so can be expected to contribute to energy shifts in the same way, leading to contributions to $\epsilon_\star$ of the form given in \eqref{epsstardist}. 

It is the final graph of Figure \ref{figure:PhotonExchange} (and its crossed counterpart) that contains the nuclear polarizability and so is not simply captured by static moments of a given nuclear distribution $\rho(\bfr) = \mel{0}{\hat\varrho(\bfr)}{0}$. For the nuclear sector this graph contributes a contribution involving a sum over nuclear states involving the off-diagonal matrix elements $|\mel{N}{\hat \varrho(\bfr)}{0}|^2$.

Explicitly, in \cite{Friar2} Friar gives the following expression for the atomic energy shift due to such a polarizability
\begin{align}
	\label{pol}
		\delta E_\text{pol} &= -\frac{4\pi}{3}(Z\alpha)^2\abs{\phi_n(0)}^2\sum_{N\neq 0}\left[ \sqrt{\frac{2m}{\omega_N}}\abs{\mel{N}{\mathbf{D}}{0}}^2 \right. \\
							&\quad \left. {}+ \frac{m}{4}\int\exd^3 x \int \exd^3 y \mel{0}{\rho(\mathbf{y})}{N}\mel{N}{\rho(\mathbf{x})}{0}\abs{\mathbf{x}-\mathbf{y}}^3\right], \notag
\end{align}
where $\mathbf{D}$ is the nucleon electric dipole operator $\mathbf{D} = \int\exd^3 x\, \mathbf{x}\rho(\bfx)$, and $\omega_N$ is the excitation energy of the intermediate nuclear state while $\phi_n(0)$ is the lepton's wavefunction at the origin. 

For the main text what is important about this calculation is that it depends on the lepton quantum numbers in precisely the same way as does the charge-radius $r_p^2$, and so can be interpreted as a shift in the value of $\epsilon_\star$. The leading (dipole) polarizability term goes as $(Z\alpha)^2\abs{\phi_n(0)}^2R^2$ and so is a contribution to the $R_1^2$ contribution of $\epsilon_\star$ in the parameterization of \eqref{eq:epsExp}.  

By comparison, the second term (and Friar moment correction) can be seen to go as relative order $(mRZ\alpha)$ and so are also contributions to $R_1^2$ in \eqref{eq:epsExp}. Contributions to $R_2^2$ in \eqref{eq:epsExp} also arise in explicit calculations, typically as relativistic kinematic corrections to lower-order terms.

\section{Solving for \texorpdfstring{$\epsilon_\star$}{epsilon\_star}}
\label{app:Astounding}

This appendix fills in the details that give the expression for $\epsilon_\star$ in situations where the energy shifts also depend logarithmically on $\epsilon_\star$. This arises in the main text when writing an expression for $\epsilon_\star$ in terms of the $2S_{1/2}-2P_{1/2}$ level shift, for example. We do so in this appendix for general $n$.

We start by writing the $nS_{1/2}-nP_{1/2}$ shift as:
\begin{equation} \label{eq:solve1}
			\widehat{\Delta E}_{nS_{1/2}-nP_{1/2}} = \frac{4m_e^3(Z\alpha)^4}{n^3}\epsilon_{\star e}^2 \left\{ \chi_n - \ln(\left(m_e\epsilon_{\star e}\right)^2) \right\},
\end{equation}
where 
\begin{align}\label{eq:defChi}
	\chi_n &:= \frac{2}{(Z\alpha)^2} - 2\ln\left(\frac{2Z\alpha}{n} \right) + \frac{12n^2-n-9}{2n^2(n+1)} - 2H_{n+1}  \notag \\
		   &\quad {}- 2\gamma + 4 + \frac{\eta^{(e)}_{n0}}{Z} - \frac{2(n^2 - 1)}{n^2} \,, 
\end{align}
and rearrange to obtain:
\begin{align}	\label{eq:solve2}
			\frac{n^3}{4m_e(Z\alpha)^4}\widehat{\Delta E}_{2S_{1/2}-2P_{1/2}} &= m_e^2\epsilon_{\star e}^2 \left\{ \chi_n - \ln(\left(m_e\epsilon_{\star e}\right)^2) \right\} \,.
\end{align}
We wish to solve this equation for $\epsilon_{\star e}$, but it has no closed-form solution. However, the solution does have a name: it is called the Lambert W-Function. In terms of this we have
\begin{equation}
			m_e^2\epsilon_{\star e}^2 = \exp[W \left( -\frac{n^3}{4m_e(Z\alpha)^4} \widehat{\Delta E}_{2S_{1/2}-2P_{1/2}} e^{-\chi_n} \right) + \chi_n]. 
\end{equation}

To turn this into something useful we use some approximate forms for $W$ in various limits. The first observation is that the argument of the W-function is order $e^{-1/(Z\alpha)^2}$ (coming from the $\chi_n$) and so is very small. Also, the energy shift in question is positive, so this argument is negative. In this limit, $W[z]$ is double-valued, and the branch of interest is the one satisfying $W[z] < -1$, denoted by $W_m[z]$. In the limit of small negative argument, 
\be
  W_m[z] \simeq -\ln\left(-\frac{1}{z} \right) - \ln\left(\ln\left(-\frac{1}{z} \right)\right) - \frac{\ln(\ln(-1/z))}{\ln(-1/z)}\dots \,, 
\ee
so that:
\begin{align}
	&W \left( -\frac{n^3}{4m_e(Z\alpha)^4} \widehat{\Delta E}_{nS_{1/2}-nP_{1/2}} e^{-\chi_n} \right) + \chi_n \notag \\
	&\qquad\simeq \ln(\frac{n^3}{4m_e(Z\alpha)^4}\widehat{\Delta E}_{nS_{1/2}-nP_{1/2}} ) \notag \\
	&\quad\qquad{} - \ln(\chi_n - \ln(\frac{n^3}{4m_e(Z\alpha)^4}\widehat{\Delta E}_{nS_{1/2}-nP_{1/2}} )) \notag \\
	&\qquad\qquad{}-\frac{(Z\alpha)^2}{2}\ln(\frac{2}{(Z\alpha)^2}) + \dots\,,
\end{align}
where the dots contain terms suppressed by order $(Z\alpha)^4$ and higher. Consequently,
\begin{align}\label{appeq:bigSol}
	m_e^2\epsilon_{\star e}^2 &\simeq  
	 \frac{n^3}{8m_e(Z\alpha)^2}\widehat{\Delta E}_{nS_{1/2}-nP_{1/2}}\left\{1 + \right. \notag \\
 &\qquad \left. (Z\alpha)^2\left[\hat\chi_n + \frac 12 \ln(\frac{n^3}{4m_e(Z\alpha)^4}\widehat{\Delta E}_{nS_{1/2}-nP_{1/2}} )\right]\right\} \notag \\
	 &\quad\qquad{}+ \order{(Z\alpha)^2\frac{n^3\widehat{\Delta E}_{nS_{1/2}-nP_{1/2}}}{4m_e}},
\end{align}
where 
\begin{align}\label{eq:defHatChi}
	\hat\chi_n &:= \frac 12 \ln(2(Z\alpha)^4/n^2) - \frac{12n^2-n-9}{4n^2(n+1)} + H_{n+1} \notag \\
	&\qquad{}+ \gamma - 2 - \frac{\eta_{n0}}{2Z} + \frac{n^2 - 1}{4n^2}\,,
\end{align} 
(defined by $\chi_n = (2/(Z\alpha)^2)(1 - \hat\chi_n)$), and the correction is down by $(Z\alpha)^4$ relative to the leading term.

Finally, plugging this into \eqref{eq:epsShift3} for the $nS_{1/2}-nP_{3/2}$ shift, we predict:
\begin{widetext}
\begin{align}\label{appeq:subP3}
   \widehat{\Delta E}_{nS_{1/2}-nP_{3/2}} &\approx \widehat{\Delta E}_{nS_{1/2}-nP_{1/2}}\left(1 + (Z\alpha)^2\left[\hat\chi_n + \frac 12 \ln(\frac{n^3}{4m_e(Z\alpha)^4}\widehat{\Delta E}_{nS_{1/2}-nP_{1/2}} )\right]\right)\notag \\
	 &\quad{}\times\left(1 + (Z\alpha)^2\left[\frac{12n^2-n-9}{4n^2(n+1)} - \frac 12\ln \left( \frac{n\widehat{\Delta E}_{nS_{1/2}-nP_{1/2}}}{2m_e} \right) - H_{n+1} - \gamma + 2 + \frac{\eta_{n0}}{2Z}\right]\right), \notag \\
 &= \widehat{\Delta E}_{nS_{1/2}-nP_{1/2}}\left(1 + (Z\alpha)^2\left[\frac{n^2- 1}{4n^2}\right] + \order{(Z\alpha)^4} \right)\,,
\end{align}
which is exactly the result \eqref{eq:goodishPredict1} used in the main text. 		
\end{widetext}


\begin{thebibliography}{99}

\bibitem{ref:LambShift}
Willis~E. Lamb and Robert~C. Retherford.
\newblock Fine {Structure} of the {Hydrogen} {Atom} by a {Microwave} {Method}.
\newblock {\em Physical Review}, 72(3):241--243, August 1947.

\bibitem{Structure}
   J.~L.~Friar and J.~W.~Negele,
  ``Theoretical and Experimental Determination of Nuclear Charge Distributions,''
  Advances in Nuclear Physics, {\bf 8} (1975) 219-376,
  
   E.~Borie,  G.~A.~Rinker,
  ``The energy levels of muonic atoms,''
  Rev.\ Mod.\ Phys.\, {\bf 54} (1982) 1 67-118 
  
   Krzysztof Pachucki,
  ``Theory of the Lamb shift in muonic hydrogen,''
  Phy.\ Rev.\ A, {\bf 53} (1996) 4, 
 
   Krzysztof Pachucki,
  ``Proton structure effects in muonic hydrogen,''
  Phys.\ Rev.\ A, {\bf 60} (1999) 5 3593--3598,
  
   Dirk Andrae,
  ``Finite nuclear charge distributions in electronic structure calculations for atoms and molecules,''
  Physics Report, {\bf 336} (2000) 6 413-525,
  
   E.~Borie,
  ``Lamb shift in muonic hydrogen,''
  Phys.~Rev.~{\bf A71} (2005) 032508.

   Michael O.~Distler, Jan C.~Bernauer, Thomas Walcher,
  ``The RMS charge radius of the proton and Zemach moments,''
  Phys.\ Lett.\ B, (2011) {\bf 696} 4 343-347,

   Carl E.~Carlson and Marc Vanderhaeghen,
  ``Higher-order proton structure corrections to the Lamb shift in muonic hydrogen,''
  Phys.\ Rev.\ A, {\bf 84} (2011) 2 020102,
  
   U.~D.~Jentschura,
  ``Lamb shift in muonic hydrogen. I. Verification and update of theoretical predictions,''
  Ann.\ Phys.\, {\bf 326} (2011) 2 500-515,

   A.~Antognini, F.~Kottmann, F.~Biraben, P.~Indelicato, F.~Nez and R.~Pohl,
  ``Theory of the 2S-2P Lamb shift and 2S hyperfine splitting in muonic hydrogen,''
  Ann.\ Phys.\, {\bf 331} (2013) 127-145,
  
   T.~P.~Gorringe and D.~W.~Hertzog,
  ``Precision Muon Physics,''
  Prog.\ Part.\ Nucl.\ Phys.\  {\bf 84}, 73 (2015)
  [arXiv:1506.01465 [hep-ex]].

  S.~G.~Karshenboim, V.~G.~Ivanov and E.~Y.~Korzinin,
  ``Relativistic recoil corrections to the electron-vacuum-polarization contribution in light muonic atoms,''
  Phys.\ Rev.\ A {\bf 85}, 032509 (2012)
  doi:10.1103/PhysRevA.85.032509
  [1112.2739 [physics.atom-ph]].

\bibitem{ref:exp:Walter}
H.~K. Walter et~al.
\newblock {Test of quantum-electrodynamical corrections in muonic atoms}.
\newblock {\em Phys. Lett.}, 40B:197--199, 1972.

\bibitem{ref:exp:Dixit1}
M.~S. Dixit, H.~L. Anderson, C.~K. Hargrove, R.~J. Mckee, D.~Kessler, H.~Mes,
  and Albert~C. Thompson.
\newblock {Experimental test of the theory of muonic atoms}.
\newblock {\em Phys. Rev. Lett.}, 27:878--881, 1971.

\bibitem{ref:th:Sundaresan}
M.~K. Sundaresan and P.~J.~S. Watson.
\newblock Higher-order vacuum polarization corrections in muonic atoms.
\newblock {\em Phys. Rev. Lett.}, 29:15--18, Jul 1972.

\bibitem{ref:th:Blomqvist}
J.~Blomqvist.
\newblock Vacuum polarization in exotic atoms.
\newblock {\em Nuclear Physics B}, 48(1):95 -- 103, 1972.

\bibitem{ref:th:WiletsRinker}
L.~{Wilets} and G.~A. {Rinker}, Jr.
\newblock {Estimate of the (Z{$\alpha$})$^{2}${$\alpha$}$^{2}$ Vacuum
  Polarization Term in Muonic Pb}.
\newblock {\em Physical Review Letters}, 34:339--341, February 1975.

\bibitem{ref:th:RinkerWilets}
G.~A. Rinker and L.~Wilets.
\newblock {Vacuum polarization in high-z, finite-size nuclei}.
\newblock {\em Phys. Rev. Lett.}, 31:1559--1562, 1973.

\bibitem{ref:exp:Dixit2}
M.~S. Dixit, A.~L. Carter, E.~P. Hincks, D.~Kessler, J.~S. Wadden, C.~K.
  Hargrove, R.~J. McKee, H.~Mes, and H.~L. Anderson.
\newblock New muonic-atom test of vacuum polarization.
\newblock {\em Phys. Rev. Lett.}, 35:1633--1635, Dec 1975.

\bibitem{ref:exp:Kaeser}
K~Kaeser, B~Robert-Tissot, L~A Schaller, L~Schellenberg, and H~Schneuwly.
\newblock {PRECISION} {TEST} {OF} {VACUUM} {POLARIZATTON} {IN} {HEAVY} {MUONIC}
  {ATOMS} t.
\newblock page~20.

\bibitem{ref:resolved} 
  J.~L.~Vuilleumier, W.~Dey, R.~Engfer, H.~Schneuwly, H.~K.~Walter and A.~Zehnder,
  ``Test of electron Screening and Vacuum Polarization in Heavy Muonic Atoms,''
  Z.\ Phys.\ A {\bf 278}, 109 (1976).
  doi:10.1007/BF01437763

\bibitem{NewExp}
   A.~Antognini, et.\ al.\ ,
  ``Illuminating the proton radius conundrum: the $\mu$He+ Lamb shift,''
  Can.\ J.\ Phys.\ {\bf 89(1)} (2011) 47-57,
 
   Parthey, C.~G. and Matveev, A. and Alnis, J. and Bernhardt, B. and Beyer, A. and Holzwarth, R. and Maistrou, A. and Pohl, R. and Predehl, K. and Udem, T. and Wilken, T. and Kolachevsky, N. and Abgrall, M. and Rovera, D. and Salomon, C. and Laurent, P. and H{\"a}nsch, T.~W.,
  "Improved Measurement of the Hydrogen 1S-2S Transition Frequency,"
  Phys.\ Rev.\ Lett.\ {\bf 107} (2011) 203001
  arXiv:1107.3101 [physics.atom-ph].

   N.~T.~Amaro, F.~D.~Antognini,
  ``The Lamb-shift experiment in Muonic helium,''
  Hyperfine Interact (2012) 212: 195.

   A.~Antognini, et al.,
  ``Experiments towards resolving the proton charge radius puzzle,''
  arXiv:1509.03235 [physics.atom-ph],
 
  R.~Pohl {\it et al.},
  ``Laser Spectroscopy of Muonic Atoms and Ions,''
  JPS Conf.\ Proc.\  {\bf 18}, 011021 (2017)
  doi:10.7566/JPSCP.18.011021
  [arXiv:1609.03440 [physics.atom-ph]].


\bibitem{QEDShifts}
See e.g.,
   Peter J.~Mohr, David B.~Newell and Barry N.~Taylor, 
   ``CODATA recommended values of the fundamental physical constants: 2014,'' 
  Rev.~Mod.~Phys.~{\bf 88} (2016) 035009-1.


\bibitem{PRNew}
   V.~Barger, C.~W.~Chiang, W.~Y.~Keung and D.~Marfatia,
  ``Proton size anomaly,''
  Phys.\ Rev.\ Lett.\  {\bf 106} (2011) 153001
  [arXiv:1011.3519 [hep-ph]];

   D.~Tucker-Smith and I.~Yavin,
  ``Muonic hydrogen and MeV forces,''
  Phys.\ Rev.\ D {\bf 83} (2011) 101702
  [arXiv:1011.4922 [hep-ph]];
  
   B.~Batell, D.~McKeen and M.~Pospelov,
  ``New Parity-Violating Muonic Forces and the Proton Charge Radius,''
  Phys.\ Rev.\ Lett.\  {\bf 107} (2011) 011803
  [arXiv:1103.0721 [hep-ph]];
  
   A.~De R\'ujula,
 ``QED confronts the radius of the proton,''
  Phys.\ Lett.\ B {\bf 697} (2011) 26-31,
   
   C.~E.~Carlson and B.~C.~Rislow,
  ``New Physics and the Proton Radius Problem,''
  Phys.\ Rev.\ D {\bf 86} (2012) 035013
  [arXiv:1206.3587 [hep-ph]].
   
   R.~Onofrio,
  ``Proton radius puzzle and quantum gravity at the Fermi scale,''
  EPL {\bf104} (2013) 20002,
  
   Li-Bang Wang and Wei-Tou Ni,
  ``Proton Radius Puzzle and Large Extra Dimensions,''
  Mod.\ Phys.\ Lett.\ A {\bf 28} (2013) 1350094
  
   S.~G.~Karshenboim, D.~McKeen and M.~Pospelov,
  ``Constraints on muon-specific dark forces,''
  Phys.\ Rev.\ D {\bf 90} (2014) no.7,  073004
  Addendum: [Phys.\ Rev.\ D {\bf 90} (2014) no.7,  079905]
  [arXiv:1401.6154 [hep-ph]].
 
   D.~Robson,
  ``Solution to the proton radius puzzle,''
  Int.\ J.\ Mod.\ Phys.\ E {\bf 23} (2015) no.12,  1450090
  [arXiv:1305.4552 [nucl-th]].
 
   P.~Brax and C.~Burrage,
  ``Explaining the proton radius puzzle with disformal scalars,''
  Phys. Rev. D {\bf 91} (2015) 043515,
  


\bibitem{ProtonRadius}

   K.~A.~Woodle {\it et al.},
  ``Measurement of the Lamb Shift in the $N=2$ State of Muonium,''
  Phys.\ Rev.\ A {\bf 41} (1990) 93.

   R.~Pohl, et al.\ ,
  ``The size of the proton,''
  Nature {\bf 466} (2010) 213-216,
  
   R.~Pohl, R.~Gilman, G.~A.~Miller and K.~Pachucki,
  ``Muonic hydrogen and the proton radius puzzle,''
  Ann.\ Rev.\ Nucl.\ Part.\ Sci.\  {\bf 63} (2013) 175
  [arXiv:1301.0905 [physics.atom-ph]];
  
   A.~Antognini, et al.\ ,
   ``Proton Structure from the Measurement of 2S-2P Transition Frequencies of Muonic Hydrogen,''
   Science {\bf 339} (2013) 417-420,
 
   R.~ Pohl, R.~Gilman, G.~A.~Miller and K.~Pachucki,
  ``Muonic Hydrogen and the Proton Radius Puzzle,''
  Annu.\ Rev.\  Nucl.\ Part.\ Sci.\ {\bf 63} (2013) 175-204,
  
   C.~E.~Carlson,
  ``The Proton Radius Puzzle,''
  Prog.\ Part.\ Nucl.\ Phys.\  {\bf 82} (2015) 59
  [arXiv:1502.05314 [hep-ph]].
  
   Randolf Pohl,
  ``Laser Spectroscopy of Muonic Hydrogen and the Puzzling Proton,''
  J. Phys. Soc. Jpn {\bf 85} (2016) 091003,
 
   E.~J.~Downie,
  ``The Proton Radius Puzzle,''
  EPJ Web of Conferences {\bf 113} (2016) 05021,
 
   J.~J.~Krauth et al.\ ,
  ``The proton radius puzzle,''
  arXiv:1706.00696 [physics.atom-ph]
 
\bibitem{QEDEFT}
   W.~E.~Caswell and G.~P.~Lepage,
  ``Effective Lagrangians for Bound State Problems in QED, QCD, and Other Field Theories,''
  Phys.\ Lett.\  {\bf 167B} (1986) 437.

   C.~Peset and A.~Pineda,
  ``The Lamb shift in muonic hydrogen and the proton radius from effective field theories,''
  Eur.\ Phys.\ J.\ A {\bf 51} (2015) no.12,  156
  [arXiv:1508.01948 [hep-ph]].
  
   A.\ A.\ Krutov, A.\ P.\ Martynenko, G.\ A.\ Martynenko, et al.\ ,
  J.\ Exp.\ Theor.\ Phys.\ {\bf 120} (2015) 73,



\bibitem{ProtonRadiusEFT}

   A.~Pineda,
  ``The Chiral structure of the Lamb shift and the definition of the proton radius,''
  Phys.\ Rev.\ C {\bf 71} (2005) 065205
  [hep-ph/0412142];
 	  
   R.~J.~Hill and G.~Paz,
  ``Model independent extraction of the proton charge radius from electron scattering,''
  Phys.\ Rev.\ D {\bf 82} (2010) 113005
  [arXiv:1008.4619 [hep-ph]];

    C.~Peset and A.~Pineda,
  ``Model-independent determination of the Lamb shift in muonic hydrogen and the proton radius,''
  Eur.\ Phys.\ J.\ A {\bf 51} (2015) no.3,  32
  [arXiv:1403.3408 [hep-ph]];
   
   C.~Peset and A.~Pineda,
  ``The Lamb shift in muonic hydrogen and the proton radius from effective field theories,''
  Eur.\ Phys.\ J.\ A {\bf 51} (2015) no.12,  156
  [arXiv:1508.01948 [hep-ph]];
  
  M.~Horbatsch and E.~A.~Hessels,
  ``Evaluation of the strength of electron-proton scattering data for determining the proton charge radius,''
  Phys.\ Rev.\ C {\bf 93} (2016) no.1,  015204
  [arXiv:1509.05644 [nucl-ex]].
  
   G.~Lee, J.~R.~Arrington and R.~J.~Hill,
  ``Extraction of the proton radius from electron-proton scattering data,''
  Phys. Rev. D {\bf 92} (2015) 013013,
 
    
\bibitem{PPEFT}
   C.~P.~Burgess, P.~Hayman, M.~Williams and L.~Zalavari,
  ``Point-Particle Effective Field Theory I: Classical Renormalization and the Inverse-Square Potential,''
  JHEP {\bf 1704} (2017) 106
  [arXiv:1612.07313 [hep-ph]].
  
\bibitem{KG}
   C.~P.~Burgess, P.~Hayman, M.~Rummel, M.~Williams and L.~Zalavari,
  ``Point-Particle Effective Field Theory II: Relativistic Effects and Coulomb/Inverse-Square Competition,''
  JHEP {\bf 1707} (2017) 072
  [arXiv:1612.07334 [hep-ph]].

\bibitem{Dirac}
   C.~P.~Burgess, P.~Hayman, M.~Rummel and L.~Zalavari,
  ``Point-Particle Effective Field Theory III: Relativistic Fermions and the Dirac Equation,''
  arXiv:1706.01063 [hep-ph].

\bibitem{Dirac2}
C.~P.~Burgess, P.~Hayman, M.~Rummel,
``Point-Particle Effective Field Theory and Subleading Finite-Size Effects in Nuclear Spin-0 Atoms''
In preparation.

\bibitem{Eides} 
   M.~I.~Eides, H.~Grotch and V.~A.~Shelyuto,
  ``Theory of light hydrogen - like atoms,''
  Phys.\ Rept.\  {\bf 342}, 63 (2001)
  [hep-ph/0002158].
  
\bibitem{Zemach}
   A.C.~Zemach,
  ``Proton Structure and the Hyperfine Shift in Hydrogen,''
  Phys.~Rev. {\bf 104} (1956) 1771.
 
\bibitem{Friar}
   J.~L.~Friar,
  ``Nuclear Finite Size Effects in Light Muonic Atoms,''
  Annals Phys.\  {\bf 122} (1979) 151.

\bibitem{Nickel}
   Bernie Nickel, 
  ``Nuclear size effects on hydrogenic atom energies: a semi-analytic formulation'',
  J.\ Phys.\ B: At.\ Mol.\ Opt.\ Phys.\  {\bf 46} 2013, 015001\ ,

\bibitem{NASA}
   R.T.~Deck, J.G.~Amar and G.~Fralick
  ``Nuclear size corrections to the energy levels of single-electron and -muon atoms,''
  Journal of Physics B: At.~Mol.~Opt.~Phys. {\bf 38} (2005) 2173-2186.

\bibitem{Ji} 
   C.~Ji, N.~Nevo Dinur, S.~Bacca and N.~Barnea,
  ``Nuclear Polarization Corrections to the $\mu^4$He$^+$ Lamb Shift,''
  Phys.\ Rev.\ Lett.\  {\bf 111}, 143402 (2013)
  [arXiv:1307.6577 [nucl-th]].

\bibitem{hfs}
S.~G. Karshenboim and V.~G. Ivanov.
\newblock {The European Physical Journal D - Atomic, Molecular, Optical and
  Plasma Physics}, 19(1):13--23, Apr 2002.

\bibitem{Lamberg}
See e.g.,
M.~Weitz, A.~Huber, F.~Schmidt-Kaler, D.~Leibfried, W.~Vassen, C.~Zimmermann,
  K.~Pachucki, T.~W. H\"ansch, L.~Julien, and F.~Biraben.
\newblock {Phys. Rev. A}, 52:2664--2681, Oct 1995.

Savely~G Karshenboim.
\newblock {Physics Reports}, 422(1):1--63, 2005.

\bibitem{fsQED}
Eides M.I., Grotch H., Shelyuto V.A. (2007) Lamb Shift in Light Muonic Atoms. In: Theory of Light Hydrogenic Bound States. Springer Tracts in Modern Physics, vol 222. Springer, Berlin, Heidelberg

\bibitem{fsQEDelec}
Eides M.I., Grotch H., Shelyuto V.A. (2007) Nuclear Size and Structure Corrections. In: Theory of Light Hydrogenic Bound States. Springer Tracts in Modern Physics, vol 222. Springer, Berlin, Heidelberg

\bibitem{Friar2}
  J.~L.~Friar,
  ``Nuclear Polarization Corrections to $\mu-d$ Atoms in Zero-Range Approximation,''
  Phys.\ Rev.\ C {\bf 88} (2013) no.3,  034003
  [arXiv:1306.3269 [nucl-th]].

\bibitem{Ji2018}
  C.~Ji, S.~Bacca, N.~Barnea, O.~J.~Hernandez and N.~Nevo-Dinur,
  ``Ab initio calculation of nuclear structure corrections in muonic atoms,''
  arXiv:1806.03101 [nucl-th].
  
\bibitem{3Photon}
K.~Pachucki, V.~Patk\'{o}\v{s} and V.~A.~Yerokhin,
  ``Three-photon exchange nuclear structure correction in hydrogenic systems,''
  Phys.\ Rev.\ A {\bf 97} (2018) no.6,  062511
  [arXiv:1803.10313 [physics.atom-ph]].

\bibitem{QEDPPe}
  Michael~I. Eides, Howard Grotch, and Valery~A. Shelyuto.
  \newblock {\em Nuclear Size and Structure Corrections}, pages 109--130.
  \newblock Springer Berlin Heidelberg, Berlin, Heidelberg, 2007.

\bibitem{QEDPPmu}
  Michael~I. Eides, Howard Grotch, and Valery~A. Shelyuto.
  \newblock {\em Lamb Shift in Light Muonic Atoms}, pages 131--159.
  \newblock Springer Berlin Heidelberg, Berlin, Heidelberg, 2007.
  See also:
  K.~Pachucki,
  ``Theory of the Lamb shift in muonic hydrogen,''
  Phys.\ Rev.\ A {\bf 53} (1996) 2092.
  doi:10.1103/PhysRevA.53.2092

\bibitem{PachuckiPolRev}
  K.~Pachucki and A.~M.~Moro,
  ``Nuclear polarizability of helium isotopes in atomic transitions,''
  Phys.\ Rev.\ A {\bf 75} (2007) 032521
  doi:10.1103/PhysRevA.75.032521
  [nucl-th/0612065].

\bibitem{BigNuc}
  G.~Audi, F.~G.~Kondev, M.~Wang, W.~J.~Huang and S.~Naimi,
  ``The NUBASE2016 evaluation of nuclear properties,''
  Chin.\ Phys.\ C {\bf 41} (2017) no.3,  030001.
  doi:10.1088/1674-1137/41/3/030001

\bibitem{HeStoSExp}
M.~{Herrmann}, M.~{Haas}, U.~D. {Jentschura}, F.~{Kottmann}, D.~{Leibfried},
  G.~{Saathoff}, C.~{Gohle}, A.~{Ozawa}, V.~{Batteiger}, S.~{Kn{\"u}nz},
  N.~{Kolachevsky}, H.~A. {Sch{\"u}ssler}, T.~W. {H{\"a}nsch}, and T.~{Udem}.
\newblock {Feasibility of coherent xuv spectroscopy on the $1S-2S$ transition in
  singly ionized helium}.
\newblock {\em Physical Review A}, 79(5):052505, May 2009.

\bibitem{HeLambExp}
A.~{van Wijngaarden}, F.~{Holuj}, and G.~W. {Drake}.
\newblock {Lamb shift in He$^{+}$: Resolution of a discrepancy between theory
  and experiment}.
\newblock {\em Physical Review A}, 63:012505, January 2001.

\bibitem{Yerokhin}
V.~A. {Yerokhin} and V.~M. {Shabaev}.
\newblock {Lamb Shift of n = 1 and n = 2 States of Hydrogen-like Atoms, 1 $\le$
  Z $\le$ 110}.
\newblock {\em Journal of Physical and Chemical Reference Data}, 44(3):033103,
  September 2015.

\end{thebibliography}
\end{document}